\documentclass[12pt,number]{elsarticle}
\usepackage[latin9]{inputenc}
\usepackage{float}
\usepackage{units}
\usepackage{siunitx}
\usepackage{hyperref}
\hypersetup{pdftitle={Phase-Field Models for Fatigue Crack Growth},
	    	pdfauthor={Ata Mesgarnejad, Northeastern University},
	    	colorlinks,
	    	pdfcreator={pdflatex},
	    	unicode=false,
	    	pdftoolbar=false,
	    	pdfmenubar=true,
	   		pdffitwindow=true,
	   		pdfnewwindow=true,
	    	linkcolor=red,
	    	citecolor=red,
	    	filecolor=black,
	    	urlcolor=red,
	    	}
\usepackage{mathrsfs}
\usepackage{amsthm}
\usepackage{amsmath}
\usepackage{amssymb}
\usepackage{amsxtra}
\usepackage{graphicx}
\usepackage{pbox}
\usepackage{algpseudocode}

\makeatletter
\usepackage[normalem]{ulem}

\floatstyle{ruled}

\newfloat{algorithm}{tbp}{loa}
\floatname{algorithm}{Algorithm}
\theoremstyle{plain}

\usepackage{pdfsync}

\usepackage{xargs}                      
\usepackage[pdftex,dvipsnames]{xcolor}
\usepackage[colorinlistoftodos,prependcaption,textsize=small]{todonotes}
\newcommandx{\comment}[2][1=]{\todo[linecolor=RubineRed,backgroundcolor=RubineRed!25,bordercolor=RubineRed,#1]{#2}}

\DeclareMathOperator*{\argmin}{\mathrm{argmin}\,}

\newcommand{\A}{\mathcal{A}}

\newcommand{\E}{\mathcal{E}}
\newcommand{\F}{\mathcal{F}}
\newcommand{\ub}{{u}}
\newcommand{\xb}{{x}}
\newcommand{\C}{{\mathcal{C}}}
\newcommand{\ie}{\emph{i.e., }}
\newcommand{\eg}{\emph{e.g., }}
\newcommand{\Fig}{Fig.}
\newcommand{\Figs}{Figs.}

\makeatother

\begin{document}

\title{Phase-Field Models for Fatigue Crack Growth}
\author[add1]{A.~Mesgarnejad}
\ead{a.mesgarnejad@neu.edu}

\author[add2]{A.~Imanian }
\ead{aimanian@tda-i.com}

\author[add1]{A.~Karma\corref{cor1}}
\cortext[cor1]{Corresponding author}
\ead{a.karma@neu.edu}

\address[add1]{Center for Inter-disciplinary Research on Complex Systems, Department of Physics, Northeastern University, Boston, MA. 02115, U.S.A.}
\address[add2]{Technical Data Analysis,3190 Fairview Park Drive, Suite 650, Falls Church, VA 22042}

\begin{abstract}
We introduce a class of models based on near crack tip degradation of  materials that can account for fracture growth under cyclic loads below the Griffith threshold.
We incorporate the gradual degradation due to a cyclic load through a flow equation that decreases spatially varying parameters controlling the fracture toughness in the vicinity of the crack tip, with the phase and displacement fields relaxed to an energy minimum at each time step.
Though our approach is phenomenological, it naturally reproduces the Paris law with high exponents that are characteristic of brittle fatigue crack growth. We show that the exponent decreases when the phase field dynamics is of the Ginzburg-Landau type with a relaxation time comparable to the cyclic loading period, or when degradation occurs on a scale larger than the process zone.
In addition to reproducing the Paris law, our approach can be used to model the growth of multiple cracks in arbitrarily complex geometries under varied loading conditions as illustrated by a few numerical examples in two and three dimensions. 
\end{abstract}

\maketitle

\section{Introduction}\label{sec:introduction}
More than one and a half centuries after the first systematic investigations by August W\"ohler~\cite{Wohler:1860}, fatigue fracture remains the major mode of failure of mechanical components. 
The pioneering experimental observations of Paris~\cite{Paris:1961} have offered a concrete connection between linear elastic fracture mechanics (LEFM) and fatigue crack propagation for long cracks under constant Mode-I loading.  
This is further evidenced by the large number of articles offering \emph{ad hoc} modifications of the power law~\cite{Paris:1963,Forman:1967,Walker:1970,Ogura:1985}.
In this article, we aim to present a phenomenological continuum model for fatigue crack growth within the framework of the phase-field approach to fracture. 
This model is shown to naturally reproduce the Paris law $da/dN\sim\Delta K^m$, which relates the crack growth per cycle to the range of variation $\Delta K$ of the mode-I stress intensity factor during cyclic loading. In addition, this model allows us to perform component level simulations of fatigue crack growth.

Paris-law type scaling has been widely observed experimentally both in ductile materials~\cite{Ritchie:1988,Suresh:1998} where the exponent $m\sim2\textup{--}4$ and in brittle materials~\cite{Bazant:1991,Ritchie:1999} where $m\sim10\textup{--}40$ is typically much higher.
Since its inception, many attempts have been made to explain the different observed exponents of the Paris power-law. 
Rice~\cite{Rice:1967} shows that simple continuum level analysis of fatigue crack growth based on perfect plasticity also predicts $m=2$.
Simple dimensional analysis with complete self-similarity assumption also predicts a second order Paris type power-law (\ie $m=2$), whereas higher $m>2$ exponents can also be justified using incomplete self-similarity arguments~\cite{Ritchie:2005,Barenblatt:2006}.
Classical continuum and meso-scale simulations have also been used to obtain $m$.
Bazant and Hubler~\cite{Bazant:2014} obtain $m=4$ through closed-form analysis of cyclic creep in quasi-brittle materials. 
Deshpande~\emph{et al.}~\cite{Deshpande:2002} obtain $m\sim4-8$ from meso-scale simulations of discrete dislocation dynamics and a simple straight cohesive crack under pure mode-I loading.
The interested reader can find a comprehensive review of different theoretical models of fatigue fracture in~\cite{Hosseini:2018}.

What distinguishes our class of models is  not only that the Paris law with a variable exponent emerges naturally under simple mode-I loading, but also that they enable direct numerical simulation of fatigue crack growth without a need for micro/meso-scale simulations and \emph{a priori} assumptions.
Within the phase-field framework, Amor~\emph{et al.}~\cite{Amor:2015} developed a fatigue model by introducing a dissipation energy that depends on  crack length. This model has the advantage of directly modeling the Paris law with a prescribed exponent, but is limited to describing the growth of a single crack.
Nguyen~\emph{et al.}~\cite{Nguyen:2001} developed a cohesive model of fatigue where the cohesive stiffness at the crack tip was reduced as a function of load rate.
Abdelmoula~\emph{et al.}~\cite{Abdelmoula:2010} introduced a similar cohesive law where the accumulation of the opening displacement was tracked by a counter variable, thereby allowing for the fatigue crack to propagate under any load. They further showed that a Paris power-law $m=4$ emerges naturally from their formulation for small loads. 
More recently, Alessi~\emph{et al.} extended this type of counter variable approach in a one-dimensional phenomenological phase-field model of fatigue~\cite{Alessi:2018a}.
Here we develop a class of phase field models that phenomenologically describe fatigue crack growth by degradation of the local fracture toughness as a result of cyclic loading following a qualitatively similar approach as Abdelmoula~\emph{et al.} \cite{Abdelmoula:2010} in a cohesive zone framework. 
The model has the advantage of describing the growth of a collection of interacting cracks in complex geometries under arbitrary loads, including fatigue and continuous growth below and above the Griffith threshold, respectively.  

We can justify our phenomenological assumptions by using the experimental observations of fatigue in brittle materials.
Even though dislocation processes that underlie ductile fatigue in crystalline solids are minimal or absent in brittle materials, failure under cyclic sub-critical loads has been widely reported for nominally brittle materials such as alumina, TZP, silicon nitride, and silica glasses as well as in ceramics and polymers (\eg see~\cite{Ritchie:1991} for a complete survey of fatigue in ceramics).
This crucial difference is also highlighted by examining postmortem crack surfaces.
Unlike in ductile fatigue of metals where there are distinct fractographical differences between the fracture surfaces of cracks in quasi-static monotonic failure and the surface of fatigue cracks, in most brittle solids there are few discernible features due to fatigue crack propagation~\cite{Dauskarat:1990}.
Nevertheless, fatigue fracture propagation in brittle materials can be explained by near-crack-tip phenomena which facilitate the sub-critical crack propagation.
For example, while bulk silicon is not susceptible to fatigue failure, cyclic failure was observed in micron scale single crystal silicons~\cite{Connally:1992,Kahn:2004,Alsem:2007} due to the creation of a silica layer which thickens at high stress sites and is prone to environmentally assisted corrosive cracking~\cite{Muhlstein:2001,Alsem:2008}.

Our work builds on a growing body of research aimed at expanding the phase-field approach to fracture.
Since their inception~\cite{Francfort:1998,Bourdin:2008a,Karma:2001a}, phase-field models of fracture have proven to be powerful tools in modeling a wide variety of fracture phenomena. These models have been validated by theoretical analyses \cite{Hakim:2009} and comparisons of predicted and observed crack paths in non-trivial geometries~\cite{Mesgarnejad:2015}. They have been used to reproduce complex experimental observations in brittle fracture including thin-film fracture~\cite{Mesgarnejad:2013},
thermal fracture \cite{Bourdin:2014a}, mixed mode fracture~\cite{Chen:2015b}, and dynamic fracture~\cite{Chen:2015b,Chen:2017a,lubomirsky2018universality}, and fracture in colloidal systems~\cite{Peco:2019} to give only a few examples. The approach has also been extended to ductile fracture~\cite{Mozaffari:2015,Ambati:2015,Borden:2016}.

This article is organized as follows: we first review the classical phase-field models of brittle fracture in section~\ref{sec:phase-field-brittle}. In section~\ref{sec:formulation}, we introduce two models of brittle fatigue fracture. We present our numerical simulation algorithms and strategy in section~\ref{sec:implementation}. 
We show in section~\ref{sub:paris-law} how the Paris law naturally emerges from our models. 
We then show two sets of simulations for complex crack paths in 2-D in section~\ref{sub:En-passant} and 3-D in section~\ref{sub:penny-shaped-crack}. 
Finally in section~\ref{sec:non-local} we show how extending the degradation zone beyond the fracture process zone results in smaller Paris law exponents.

\section{Phase-field models of brittle fracture}\label{sec:phase-field-brittle}

Brittle fracture is signified by the Griffith criterion that states that the crack propagates if and only if the energy release rate $G$ (\ie the derivative of the stored elastic energy with respect to crack length under constant loading) from the tip of the crack is larger than the fracture toughness of the material $G_c$. 

We define the free energy $\F$ by combining the elastic (bulk) and fracture (surface) energies for an elastic domain $\Omega\subset\mathbb{R}^{n}$ containing fracture set $\mathbb{R}^{n-1}\supset\Gamma\subset\Omega$ and due to deformation $\ub=\{u_i\}\,,i=1\dots N$ as
\begin{equation}\label{eq:energy-sharp}
	\F(\ub,\Gamma):=\int_{\Omega\setminus\Gamma} W(\ub)\,dV-\int_{\partial_N\Omega}\left(f\cdot\ub\right)\,ds+G_c\mathcal{H}^{n-1}(\Gamma)
\end{equation}
where $W(\ub)=(\C_{ijkl}\,e_{kl}(\ub)e_{ij}(\ub))/2$ is the elastic energy density, $e_{ij}(\ub)=(u_{i,j}+u_{j,i})/2$ is the linear strain (\ie symmetric gradient) of the displacement vector $\ub$, where for Lame parameters $\lambda,\mu$ we write the fourth order linear elastic constitutive tensor (\ie Hooke's law) as $\C_{ijkl}=\lambda \delta_{ij}\delta_{kl}+\mu(\delta_{ik}\delta_{jl}+\delta_{il}\delta_{kj})$. 
Furthermore, $\mathcal{H}^{n-1}(\Gamma)$ is the Hausdorff measure of the crack set \ie the aggregate length of cracks in two dimensions or aggregate surface area of cracks in three dimensions.
The classical elasticity boundary conditions are given as external forces $f$ acting on part of its boundary $\partial_N\Omega\subset\partial\Omega$ and Dirichlet condition $\ub\equiv\hat{\ub}$ on $\partial_N\Omega\subset\partial\Omega\setminus\Gamma$.

The phase-field method approximates the above energy by replacing the crack set $\Gamma$ with a phase field $\phi\in[0,1]$, where $\phi=1$ almost everywhere and converges to $\phi\approx 0$ at $\Gamma$. We can write the approximate free energy as:
\begin{equation}\label{eq:ATE}
	\F_{\xi}(\ub,\phi):=\mathcal{B}(\ub,\phi)+\mathcal{S}_{\xi}(\phi)
\end{equation}
where
\begin{align}
	\mathcal{B}_{\xi}(\ub,\phi):=\int_{\Omega} g(\phi)W(\ub)\,dV-\int_{\partial_N\Omega}\left(f_i u_i\right)\,ds\label{eq:ATE-Bulk}\\
	\mathcal{S}_{\xi}(\phi):=\frac{G_c}{4C_{\phi}}\int_{\Omega}\left(\frac{w(\phi)}{\xi}+\xi\left|\nabla\phi\right|^2\right)\,dV\label{eq:ATE-surface}
\end{align}
and where we introduce the phase-field length scale $\xi$. $g(\phi)$, $w(\phi)$ are monotonically increasing and decreasing functions of $\phi$ respectively where $g(1)=w(0)=1-g(0)=1-w(1)=1$, and $C_{\phi}=\int_0^1\sqrt{w(\phi)}\,d\phi$ is a scaling constant.
Throughout this manuscript we will use the formulation developed by Karma, Kessler and Levine (KKL)~\cite{Karma:2001a} and set $g(\phi)=4\phi^3-3\phi^4$ and $w(\phi)=1-g(\phi)$. 
This choice allows for the propagation of a single crack and prohibits the creation of any new cracks where $\phi=1$.
Rigorous mathematical results \cite{Bourdin:2008a} show that the minimizers of \eqref{eq:min-prob} converge to those of \eqref{eq:sharp-min-prob} when $\xi\rightarrow0$, and crack propagation laws traditionally used in the LEFM framework (Griffith criterion $G=G_c$ and the principle of local symmetry) have also been rigorously derived for this model \cite{Hakim:2009}. However, there is an ever-growing scientific consensus that $\xi$ represents an important material property that is needed for faithfully reproducing the experimental observations~\cite{Bourdin:2014a,Mesgarnejad:2015,Tanne:2017}.
Indeed, for this class of functionals defined by~\eqref{eq:ATE}-\eqref{eq:ATE-surface} one can obtain a physical process zone size using the maximum tensile strength and fracture toughness of the material~\cite{Pham:2013}.

We should also note that we can write \eqref{eq:ATE-surface} equivalently as~\cite{Karma:2001a}:
\begin{equation}\label{eq:KKL}
	\mathcal{S}_{\xi}(\phi):=\int_{\Omega}\left(\E_{c0} w(\phi)+\frac{\kappa}{2}\left|\nabla\phi\right|^2\right)\,dV
\end{equation}
where $\E_{c0}=G_c/4C_{\phi}\xi$ is interpreted in the KKL model as a critical energy density (stress) above which the $\phi=0$ state becomes energetically favored and $\kappa=G_c\xi/2C_{\phi}$ is the coefficient of the gradient square term that together with $\E_{c}$ determines the size of the process zone and fracture energy.


\subsection{Quasi-static evolution}\label{sub:qstatic}
In a quasi-static (\ie rate independent) setting where we assume that fracture and elasticity have much smaller time-scales than loading (and other variables), we can pose the problem of fracture as a minimization problem. We define the problem of fracture as:
\begin{equation}\label{eq:sharp-min-prob}
	(\ub^*,\Gamma^*)=\underset{\scriptstyle{\begin{array}{l}
								\ub\in\A_{u}\\
								\Gamma\in\A_{\Gamma}
								\end{array}}}{\argmin} \F(\ub,\Gamma)
\end{equation}
where $\mathcal{K}_A=\{\ub|\,\ub(x)=\hat{\ub}(x),\,\forall x\in\partial_{D}\Omega\}$ is the set of displacement that satisfies the boundary conditions (\ie the admissible displacements) and $\A_{\Gamma}=\{\Gamma|\,\Gamma(t^{+})\supset \Gamma(t^{-}),\,\forall t^{+}>t^{-}\}$ is the admissible fracture set.
The energy minimization problem assumes that the system evolves in such a manner as to minimize the free energy~\eqref{eq:energy-sharp}. 
It is easy to show that for a single connected crack \eqref{eq:min-prob} is equal to the Griffith fracture theory which states that quasi-static cracks only propagate at $G=G_c$.

This can be translated to its phase-field equivalent problem~\eqref{eq:sharp-min-prob} as:
\begin{equation}\label{eq:min-prob}
	(\ub^*,\phi^*)=\underset{\scriptstyle{\begin{array}{l}
								\ub\in\A_{u}\\
								\phi\in\A_{\phi}
								\end{array}}}{\argmin} \F_{\xi}(\ub,\phi)
\end{equation}
where $\A_{\phi}=\{\phi|\,0\leq\phi(t^{+})\leq \phi(t^{-})\,\forall t^{+}>t^{-}\}$.
We can then write the governing equations for $\ub$ and $\phi$ as Euler-Lagrange equations of \eqref{eq:ATE}:
\begin{align}
	&\frac{\partial \F_{\xi}}{\partial \ub}(\tilde{\ub})=0\quad\forall\tilde{\ub}\in\A_{u}\label{eq:gov-u}\\
	&\frac{\partial \F_{\xi}}{\partial \phi}(\tilde{\phi})=0\quad\forall\tilde{\phi}\in\A_{\phi}\label{eq:gov-phi-qstatic}
\end{align}
where $\dfrac{\partial\F}{\partial x}(\tilde{x})$ is the G\^ateaux derivative of free energy $\F$ with respect to $x$ in $\tilde{x}$ direction.

\subsection{Ginzburg-Landau dynamics}\label{sub:ginzburg-landau}

It is also possible to pose the problem of fracture as a gradient flow toward the local minimum energy basins while assuming a rate independent elasticity as~\eqref{eq:gov-u} (\ie ignoring the elastic kinetic effects). In this setting we write the evolution equation of the phase field $\phi$ in the standard Ginzburg-Landau form \cite{Hakim:2009}:

\begin{equation}
	\tau_{\phi}\frac{\partial \phi}{\partial t}=-\frac{1}{\E_{c0}}\min\left(0,\frac{\delta \F_{\xi}}{\delta \phi}\right)\label{eq:gov-phi-gl}
\end{equation}
where $\dfrac{\delta\F_{\xi}}{\delta\phi}$ is the Fr\'echet derivative of free energy $\F_{\xi}$ with respect to $\phi$.
Integrating both sides, \eqref{eq:gov-phi-gl1} can be equivalently written as
\begin{equation}
		\tau_{\phi}\int_{\Omega}\left(\frac{\partial \phi}{\partial t}\tilde{\phi}\right)dV=-\frac{1}{\E_{c0}}\frac{\partial \F_{\xi}}{\partial \phi}(\tilde{\phi})\quad\forall\tilde{\phi}\in\A_{\phi}\label{eq:gov-phi-gl1}
\end{equation}
where $\tau_{\phi}$ is the time scale associated with the evolution of the phase field $\phi$.
Not surprisingly, we can recover the quasi-static Euler-Lagrange equation~\eqref{eq:gov-phi-qstatic} when $\tau_{\phi}\rightarrow0$.

\section{Phase-field models of fatigue fracture}\label{sec:formulation}

Since the classical phase-field method essentially approximates the Griffith problem, it does not allow for fracture propagation at sub-critical loads \ie $G<G_c$ ($K_I<K_{IC}$).
Since fatigue fracture, by definition, happens at sub-critical loads, it cannot be modeled with classical phase-field formulations. 
Although there is an abundance of literature on the physical mechanisms of fatigue fracture initiation and propagation, here we aim to model fatigue on a phenomenological level as a dissipative process that lowers the effective near-tip fracture toughness.
This allows sub-critical cracks to propagate under cyclic loading without changing the monotonic fracture toughness.
The two models introduced subsequently are based on this observation. 

\subsection{Model-I: the variable critical energy model}\label{sub:variable-Ec}
In the first model we achieve sub-critical propagation by lowering near-tip fracture toughness by reducing the critical energy density $\E_{c}$ around the crack tip. 
By lowering $\E_{c}$ near the crack tip, in essence, we lower the energy barrier for propagation of the fracture phase-field. We rewrite the surface energy~\eqref{eq:KKL} with spatially varying $\E_{c}$:
\begin{align}\label{eq:energy-Ec} 
	\mathcal{S}_{\xi}(\phi;\E_{c})&=\int_{\Omega} \left(\E_{c} w(\phi)+\frac{\kappa}{2}|\nabla \phi|^2\right)\,dV
\end{align}
where $\E_{c}(t=0)=\E_{c0}$ and $\E_{c}(x)$ evolves according to
\begin{equation}\label{eq:evolution-ec}
	\frac{1}{\chi_c}\frac{\partial \E_{c}}{\partial t}=\min\left(0,-H(\E_{c}-\E_{c\min})\,w(\phi)\frac{\partial}{\partial t}\left(g(\phi)W^+(\ub)\right)\right)
\end{equation}
where $\chi_c$ is a scaling factor which controls the evolution of the critical energy, and $\E_{c\min}\geq0$ signifies the lowest energetic barrier against fracture propagation. We have set $\E_{c\min}=0$ for all the simulations presented in this article. In addition, we define the Heaviside step function:
\begin{equation}
	\mathrm{H}(x)=\left\{\begin{array}{rcl} 
						1&\quad&x>0\\
						0&\quad&x\leq0
					  \end{array}\right.
\end{equation}
Using the above formulation, one can immediately see that the critical energy density remains constant under a constant load but decreases when the load is increasing in time. The stress divergence at the crack tip, which is cut off on a scale $\xi$ in the phase-field model, naturally decreases $\E_{c}$ near the tip thereby allowing for fatigue propagation. However, as a first step in describing fatigue crack growth of brittle materials, we keep the degradation process localized to the process zone by introducing the multiplicative factor $w(\phi)$ in~\eqref{eq:evolution-ec}.
Furthermore, we define the positive energy potential as
\begin{equation}\label{eq:Wplus}
	W^{+}(\ub)=\left\{\begin{array}{rcl} 
						W(\ub)&\quad&e_{kk}(\ub)\geq0\\
						0&\quad&\mathrm{otherwise}
					  \end{array}\right.
\end{equation}
This allows the critical energy to only decrease in tensile stresses, thereby suppressing the propagation of compressive fatigue cracks.

Although the above formulation limits the propagation of compressive fatigue cracks, it does not limit inter-penetration of the created fatigue fracture faces under compressive loads.
It is straightforward to extend this formulation to account for non-interpenetration by additively splitting the elastic energy potential into volumetric and deviatoric parts \cite{Amor:2009}.
Furthermore, it is possible to define the positive energy potential~\eqref{eq:Wplus} to allow for shear degradation both in tension and compression using the same additive split introduced in~\cite{Amor:2009} .

To elucidate how this model allows for sub-critical propagation, we should note that in the elliptic approximations of brittle fracture, the fracture toughness can be calculated as
\begin{equation}\label{eq:Gc}
	G_c=2\sqrt{2\kappa \E_{c}}\int_{0}^{1}\sqrt{w(\phi)}\,d\phi=2\sqrt{2\kappa \E_{c}}C_{\phi}
\end{equation}
As a result, by decreasing the critical energy density in front of the crack, the near-tip fracture toughness (non-uniformly) decreases allowing for propagation of the fracture.
We should note, however, that the energy release rate for the crack to propagate (at process zone scale $\xi$) in the resulting non-uniform $\E_{c}$ field is non-trivial. 
Therefore, obtaining a closed-form solution for the equivalent near-tip fracture toughness may not be possible.

Reducing the critical energy density $\E_{c}$ in our formulation is very similar to the lowering of cohesive stiffness in~\cite{Nguyen:2001}. 
Indeed, as pointed out by Pham and Marigo~\cite{Pham:2013}, the process zone size $\xi$ and the critical energy density $\E_{c}$ in the phase-field formulation are analogous to the opening displacement and maximum stress in the cohesive zone formulation, respectively.

\subsubsection{Thermodynamical considerations}\label{subsub:thermo1}
It is straightforward to calculate the energy dissipation with respect to time for \eqref{eq:energy-Ec}. Taking the time derivative of the approximate free energy $\F_\xi$ and using the governing equations \eqref{eq:gov-u}-\eqref{eq:gov-phi-qstatic} for the quasi-static evolution and further using the fact that the local phenomenological degradation law \eqref{eq:evolution-ec} implies $\partial\E_{c}/\partial t\leq 0$, we obtain:
\begin{align}\label{eq:dissipation-Ec-quasi-static}
	\frac{d\F_{\xi}}{dt}(\ub,\phi,\E_{c})&=\int_{\Omega}\left(\frac{\delta \F_{\xi}}{\delta \ub_i}\frac{\partial \ub_i}{\partial t}+\frac{\delta \F_{\xi}}{\delta \phi}\frac{\partial \phi}{\partial t}+\frac{\delta \F_{\xi}}{\delta \E_{c}}\frac{\partial \E_{c}}{\partial t}\right)\,dV\nonumber\\
	&= \int_{\Omega} \left(w(\phi) \frac{\partial \E_{c}}{\partial t}\right)\,dV\leq0
\end{align}
which ensures the dissipative nature of the quasi-static algorithm taking into account that~\eqref{eq:gov-phi-qstatic} is equivalent to $\delta F_\xi/\delta \phi=0$.

Similarly, if the evolution of the phase field $\phi$ is governed by gradient dynamics as defined in~\eqref{eq:gov-phi-gl}, we can write:
\begin{align}\label{eq:dissipation-Ec}
	\frac{d\F_{\xi}}{dt}(\ub,\phi,\E_{c})&=\int_{\Omega}\left(\frac{\delta \F_{\xi}}{\delta \ub_i}\frac{\partial \ub_i}{\partial t}+\frac{\delta \F_{\xi}}{\delta \phi}\frac{\partial \phi}{\partial t}+\frac{\delta \F_{\xi}}{\delta \E_{c}}\frac{\partial \E_{c}}{\partial t}\right)\,dV\nonumber\\
	&= \int_{\Omega} \left(-\frac{1}{2\E_{c0}\tau_{\phi}}\left(\frac{\delta \F_{\xi}}{\delta \phi}\right)^2+w(\phi) \frac{\partial \E_{c}}{\partial t}\right)\,dV\leq0
\end{align}
Therefore, the proposed algorithm is dissipative.

\subsection{Model-II: \texorpdfstring{$\gamma$-model}{Lg}}\label{sub:variable-gamma}

It is also possible to reduce the spatial fracture toughness defining a variable fracture toughness ratio $\gamma$ in~\eqref{eq:ATE-surface}.
\begin{equation}\label{eq:energy-gamma}
		\mathcal{S}_{\xi}(\phi;\gamma):=\frac{G_{c}}{4C_{\phi}}\int_{\Omega}\gamma\left(\frac{w(\phi)}{\xi}+\xi\left|\nabla\phi\right|^2\right)\,dV
\end{equation}
where $\gamma(x)$ evolves as
\begin{equation}\label{eq:evolution-gamma}
	\frac{1}{\chi_{c}}\frac{\partial \gamma}{\partial t}  =\min\left(0, -\mathrm{H}(\gamma-\gamma_{\min})w(\phi)\ \frac{\partial}{\partial t} \left(g(\phi)\frac{W^+}{\E_{c0}} \right)\right)
\end{equation}
The minimum relative fracture toughness $\gamma_{\min}>0$ is introduced (in the same spirit as $\E_{c\min}$~\eqref{eq:evolution-ec}) as the lowest relative fracture toughness achievable. 
It should be noted that (if $\gamma_{\min}=0$) the limit $\gamma\rightarrow 0$ of~\eqref{eq:energy-gamma} can result in $\phi=0$ at a dense subdomain of $\Omega$ which results in an ill-defined elasticity problem.

Similar to the first model, it is worth emphasizing that unlike the classical phase-field formulation with uniform fracture toughness, the resulting fracture toughness near the crack tip defined by~\eqref{eq:energy-gamma} is non-trivially related to $G_{c}$ because 
$\gamma$ is spatially varying in the crack tip region. 

\subsubsection{Thermodynamical considerations}\label{subsub:thermo2}
Similar to section~\ref{subsub:thermo1} taking the derivative of the approximate free energy $\F_\xi$ with respect to time and taking in account the quasi-static governing equations \eqref{eq:gov-u}--\eqref{eq:gov-phi-qstatic} we can write:
\begin{align}\label{eq:dissipation-gamma-quasi-static}
	\frac{d\F_{\xi}}{dt}(\ub,\phi)&=\int_{\Omega}\left(\frac{\delta \F_{\xi}}{\delta \ub_i}\frac{\partial \ub_i}{\partial t}+\frac{\delta \F_{\xi}}{\delta \phi}\frac{\partial \phi}{\partial t}+\frac{\delta \F_{\xi}}{\delta \gamma}\frac{\partial \gamma}{\partial t}\right)\,dV\nonumber\\
	&= \int_{\Omega} \left(\frac{G_{c}}{4C_{\phi}}\left(\frac{w(\phi)}{\xi}+\xi\left|\nabla\phi\right|^2\right)\frac{\partial\gamma}{\partial t}\right)\,dV\leq0
\end{align}
once again noticing that ${\partial\gamma}/{\partial t}\leq 0$ as imposed by the phenomenological degradation law~\eqref{eq:evolution-gamma}.

The same procedure can be duplicated for the gradient dynamics:
\begin{align}\label{eq:dissipation-gamma}
	\frac{d\F_{\xi}}{dt}(\ub,\phi)&=\int_{\Omega}\left(\frac{\delta \F_{\xi}}{\delta \ub_i}\frac{\partial \ub_i}{\partial t}+\frac{\delta \F_{\xi}}{\delta \phi}\frac{\partial \phi}{\partial t}+\frac{\delta \F_{\xi}}{\delta \gamma}\frac{\partial \gamma}{\partial t}\right)\,dV\nonumber\\
	&= \int_{\Omega} \left(-\frac{1}{\E_{c0}\tau_{\phi}}\left(\frac{\delta \F_{\xi}}{\delta \phi}\right)^2+\frac{G_{c}}{4C_{\phi}}\left(\frac{w(\phi)}{\xi}+\xi\left|\nabla\phi\right|^2\right)\frac{\partial\gamma}{\partial t}\right)\,dV\leq0
\end{align}

\section{Numerical implementation}\label{sec:implementation}
\subsection{Dimensionless parameters}\label{sub:dimensionless}

In all sections that follow, we define dimensionless spatial coordinates using one of the sample dimensions $\bar{\xb}=\xb/\xi$ and displacements by $\bar{\ub}=\ub/h_0$ where we define $h_0=\sqrt{G_{c}\,\xi/E}=K_{IC}\sqrt{\xi}/E$. We can write the dimensionless free energy form of~\eqref{eq:ATE} as:

\begin{align}
	\bar{\F}_{\xi}(\ub,\phi):=\frac{{\F}_{\xi}(\ub,\phi)}{G_{c}\xi^{n-1}}=&\frac{1}{2}\int_{\bar{\Omega}} g(\phi)\left(\bar{\C}_{ijkl} \bar{e}_{kl}(\ub)\bar{e}_{ij}(\ub)\right)\,d\bar{V}\nonumber\\-&\int_{\partial_N\bar{\Omega}}\left(\bar{f}_i\bar{u}_i\right)\,d\bar{s}+\frac{1}{4C_{\phi}}\int_{\bar{\Omega}}\left(w(\phi)+\left|\nabla\phi\right|^2\right)\,d\bar{V}\label{eq:ATE-nD}
\end{align}
where $\bar{\C}_{ijkl}=\C_{ijkl}/E$, and $\bar{f}_i=f_i\sqrt{\xi/G_{c0}E}$. Furthermore, for the dynamic phase-field evolution~\eqref{eq:gov-phi-gl1}, we use $\tau_{\phi}$ to adimensionalize time \ie $\bar{t}=t/\tau_{\phi}$.

\subsection{Numerical algorithms}\label{sub:algorithms}
In this article, the spatial discretization is done using the Galerkin finite element method. The critical energy field $\E_{c}$ and the fracture toughness ratio parameter $\gamma$ are discretized as discontinuous constants across each element (\ie DG-0), and components of $\ub$ and $\phi$ are discretized using Lagrange basis functions.
\subsubsection{Alternate minimum algorithm for quasi-static evolution}\label{subsub:alt-min-algo}
Our minimization strategy for~\eqref{eq:ATE} is now classical~\citep{Bourdin:2008a}. 
At each time step, it is achieved by alternating minimizations with respect to $u$ and $\phi$ until convergence, leveraging the separate convexity of the regularized energy with respect to each field. 
As stated before, the solution for displacement $\ub$ and phase field $\phi$ is calculated by solving their respective Euler-Lagrange equations~\eqref{eq:gov-u}-\eqref{eq:gov-phi-qstatic}.
The first step to obtain $\ub$ using the alternate minimization algorithm is a simple convex problem implemented by solving the elastic equilibrium. 
We use a bounded reduced space Newton minimization scheme~\citep{Benson:2006} for the second step to solve for $\phi$ .
The phase-field method requires the spatial resolution of discretization to be smaller than the process zone length $\xi$.

As presented in algorithm~\ref{algo:alt-min}, the variables of the fatigue models $\E_{c}$ or $\gamma$ are only updated outside the alternate minimization loop after the alternate minimization algorithm converges at each time step \ie
\begin{equation}
	{\E_{c}}_{n+1} = \max \left({\E_{c\min}},\min\left({\E_{c}}_{n},{\E_{c}}_{n} - \chi_c\left({g(\phi_{n+1})W^+(u_{n+1})-g(\phi_{n})W^+(u_n)}\right)\right)\right)
	\label{iter1}
\end{equation}
or
\begin{equation}
	{\gamma}_{n+1} = \max \left({\gamma_{\min}},\min\left({\gamma}_{n},{\gamma}_{n} - \chi_c\left({g(\phi_{n+1})W^+(u_{n+1})-g(\phi_{n})W^+(u_n)}\right)\right)\right)  
	\label{iter2}
\end{equation}
The resulting problems are often very large and require the use of a parallel programming paradigm and the complex numerical tools therein. Our implementation~\cite{Mesgarnejad:2015} relies on the distributed data structures provided by \texttt{libMesh}~\citep{libmesh} and linear algebra provided by \texttt{PETSc}~\citep{petsc-efficient,petsc-user-ref}.

\begin{algorithm}[H]
\begin{algorithmic}[1]
\State{Set $\phi_{0}=1$.}
\State{Let $\delta_{altmin}$ be a given tolerance parameters.}
\For{$n=0$ to $N$}
	\State{$\phi^{0}\longleftarrow\phi_{n-1}$}
	\While{$\left| \phi^j-\phi^{j-1}\right|_{L^\infty} \ge \delta_{altmin}$}
		\State{$u^{j+1}\longleftarrow\underset{u\in\A_{u}}{\argmin}\F_{\xi}(\ub,\phi^{j})$}
		\State{$\phi^{j+1}\longleftarrow\underset{\phi\in\A_{\phi}}{\argmin}\F_{\xi}(\ub^{j+1},\phi)$}
		\State{$j\longleftarrow j+1$}
	\EndWhile
	\State{$\ub_{n}\longleftarrow \ub^{j}$}
	\State{$\phi_{n}\longleftarrow\phi^{j}$}
	\State{Update $\E_{c}$ or $\gamma$ from~\eqref{eq:evolution-ec} or~\eqref{eq:evolution-gamma}.}
\EndFor
\end{algorithmic}
\caption{the alternate minimization algorithm.}
\label{algo:alt-min}
\end{algorithm}

\subsubsection{Dynamic algorithm}\label{subsub:dynamic-algo}
The algorithm~\ref{algo:dynamic} for dynamic evolution is much simpler and only ensures elastic equilibrium at each time step. We have implemented an implicit forward Euler time stepping algorithm for $\phi$ where at each time step we solve:
\begin{equation}\label{eq:dynamic-phi}
	\int_{\Omega} \left(\frac{\phi_{t}-\phi_{t-1}}{\delta t}\right)\tilde{\phi}\,dV=-\frac{\partial \F_{\xi}}{\partial \phi}(\tilde{\phi})\quad\forall\tilde{\phi}\in\A_{\phi}
\end{equation}
using the bounded reduced space Newton minimization scheme implemented in PETSc~\cite{petsc-user-ref}.
\begin{algorithm}[H]
\begin{algorithmic}[1]
\State{Set $\phi_{0}=1$.}
\For{$n=0$ to $N$}
	\State{Update $\phi_n$ by solving implicit time integration~\eqref{eq:dynamic-phi}}
	\State{$u_n\longleftarrow\underset{u\in\A_{u}}{\argmin}\F_{\xi}(\ub,\phi_{n})$}
	\State{Update $\E_{c}$ or $\gamma$ from~\eqref{eq:evolution-ec} or~\eqref{eq:evolution-gamma}.}
\EndFor
\end{algorithmic}
\caption{the dynamic algorithm.}
\label{algo:dynamic}
\end{algorithm}

\section{Numerical simulations}\label{sec:simulations}
\subsection{Paris law}\label{sub:paris-law}
\begin{figure}[htb!]
	\centering
	\includegraphics[width=.65\textwidth]{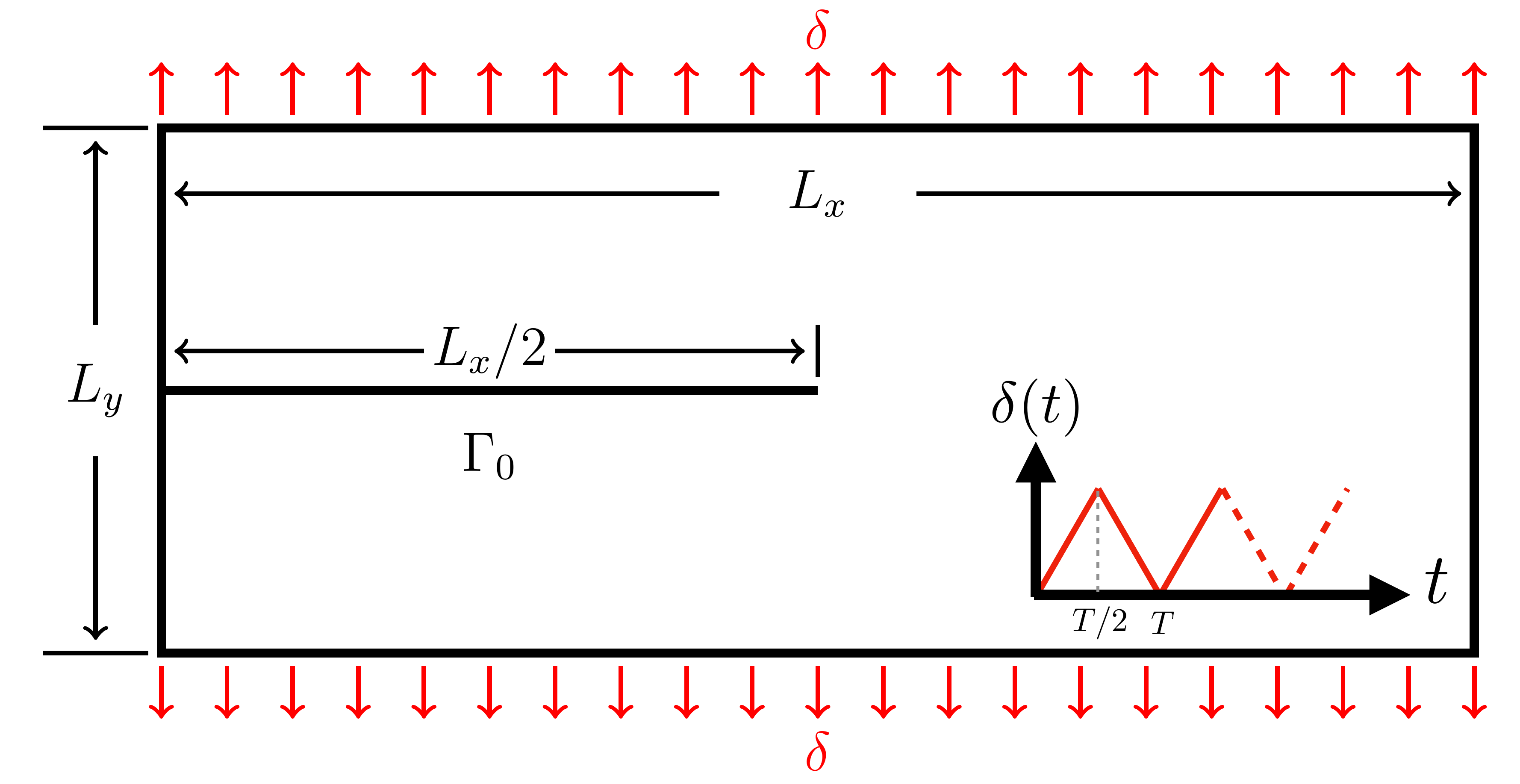}
	\caption{Schematics of the mode-I crack geometry used for the fatigue crack growth simulations with the loading profile is depicted in the inlet.}
	\label{fig:schematics-favorite-geom}
\end{figure}

The first validation on any continuum fatigue model is to assess its ability to simulate the Paris law. This law states that the increase in the length of a fatigue crack per unit cycle relates to the range of applied stress intensity factors as a power-law \ie $da/dN\sim(\Delta K)^n$ where $a$ is the crack length and $N$ is the number of cycles.
For this purpose, we set up a simple problem illustrated in \Fig~\ref{fig:schematics-favorite-geom} for a  rectangular domain $\Omega=(-L_x/2,L_x/2)\times(-L_y/2,L_y/2)$ containing an initial crack $\Gamma_0=(-L_x/2,0)\times{0}$.
We impose an oscillatory vertical displacement on the top and bottom  $u_y(\pm L_y)=\delta(t)$ where $\delta(t)$ is oscillating in a sawtooth fashion by linearly increasing between $0$ and $\delta_{\max}$ when $t$ increases from $nT$ to $nT+T/2$ and linearly decreasing from $\delta_{\max}$ to $0$ when $t$ increases from $nT+T/2$ to $(n+1)T$, where the integer $n$ is the cycle number and $T$ is the period. 
Additionaly, we set the horizontal displacement $u_x(\pm L_y)=0$.
All the simulations in this section are done for $L_y=50\xi$ and $L_x=200\xi$.

The energy release rate from the tip of a straight crack in the geometry shown in \Fig~\ref{fig:schematics-favorite-geom} can be approximated by $G\simeq2E\delta^2/(1-\nu^2) L_y$. 
Using this formula, we performed the numerical simulation at different $\delta_{\max}$ corresponding to different $G/G_c$ ($K_I/K_{IC}$) ratios where we set $\nu=1/3$. We note that here $\Delta K/K_{IC}=K_I/K_{IC}$ since the material is completely unloaded at each cycle.

\begin{figure}[htb!]
	\centering
	\includegraphics[width=.75\textwidth]{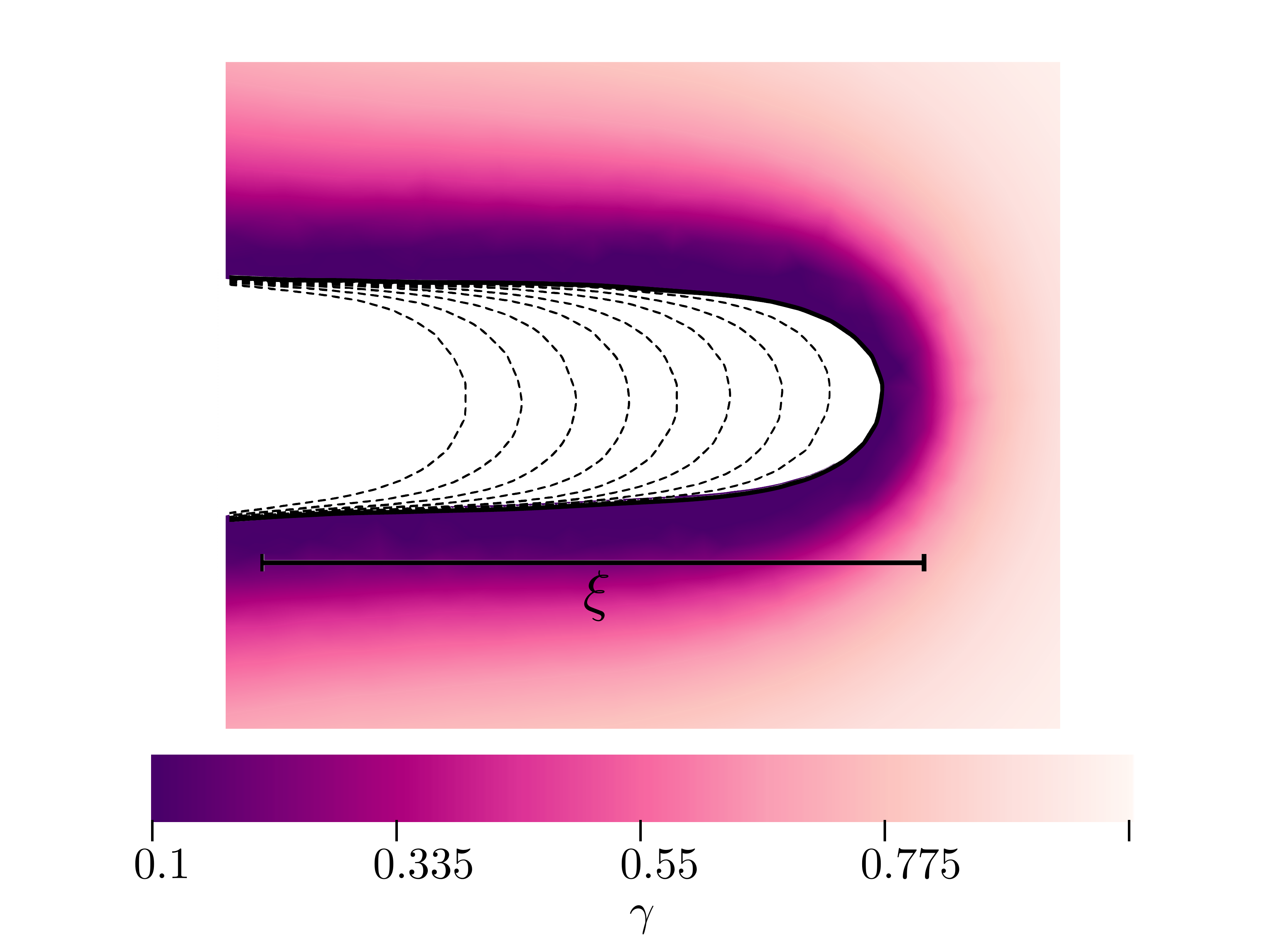}
	\caption{Crack tip (\ie contours of $\phi=0.5$) propagation under mode-I cyclic loading (see figure~\ref{fig:schematics-favorite-geom}) using the $\gamma$-model for $L_x/L_y=4$, $\nu=1/3$, $\chi_{c}=1$, and $K_I/K_{IC}=3/4$  plotted for 40 cycles at 5 cycle intervals. The color map showing the fracture toughness ratio field $\gamma$ for the last cycle outside the crack (\ie $\phi\geq0.5$) was superimposed.}
	\label{fig:local-snapshot}
\end{figure}

\Fig~\ref{fig:local-snapshot} shows the evolution of the crack ($\phi=0.5$ contour line) over 40 cycles (after the initial transient regime) as well as the spatial distribution of $\gamma$ at the last step. 
The translational invariance of the geometry and the phase-field formulation result in a steady state propagation of the initial crack with $da/dN$ approaching a constant after a transient regime lasting a few cycles.

\begin{figure}[htb!]
	\centering
	\includegraphics[width=.9\textwidth]{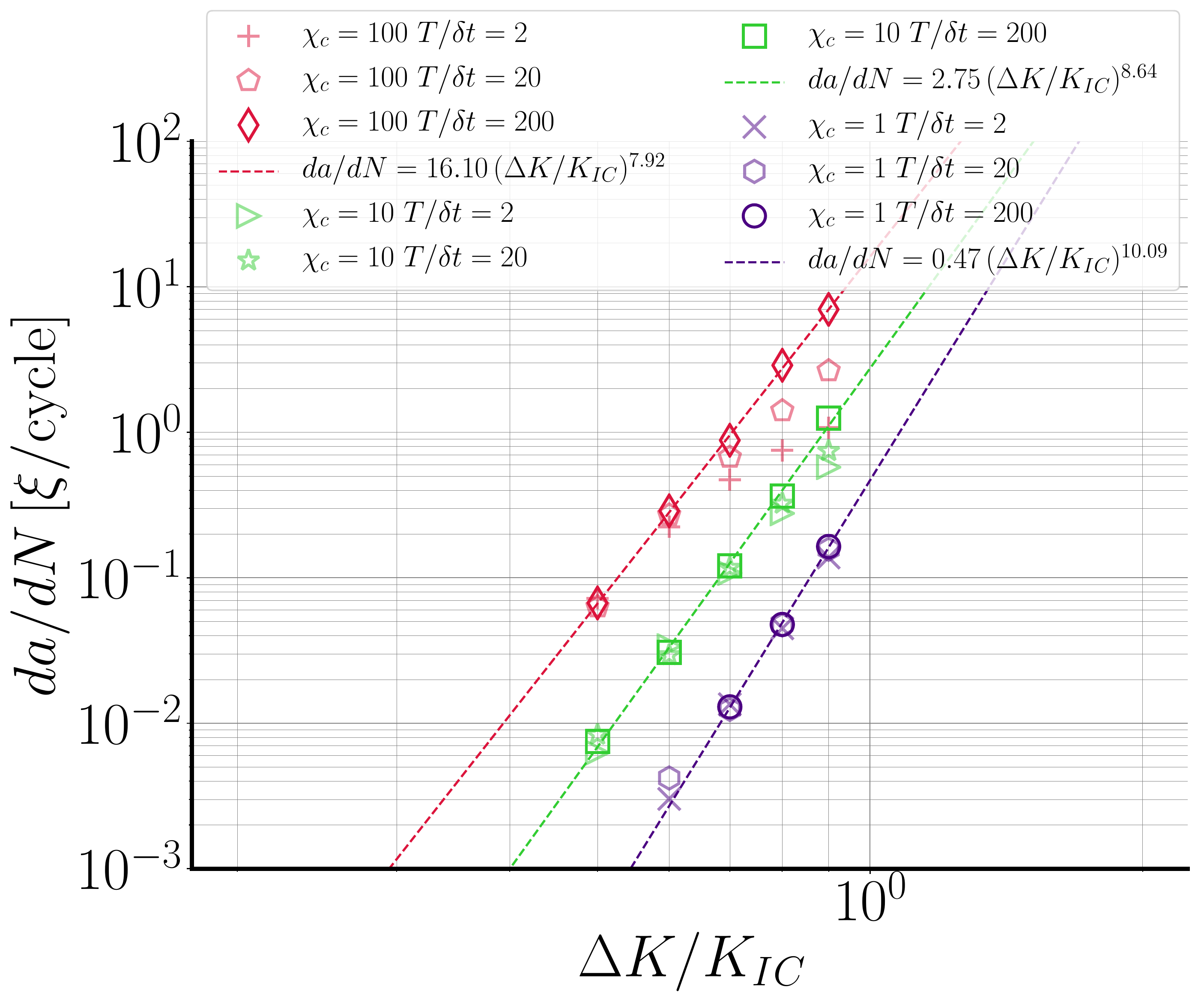}
	\caption{Emergent Paris law in quasi-static simulations of single mode-I crack for the variable critical energy model~\ref{sub:variable-Ec} showing the convergence for different number of time steps per cycle. }
	\label{fig:Paris-Ec-qstatic}
\end{figure}

\begin{figure}[htb!]
	\centering
	\includegraphics[width=.9\textwidth]{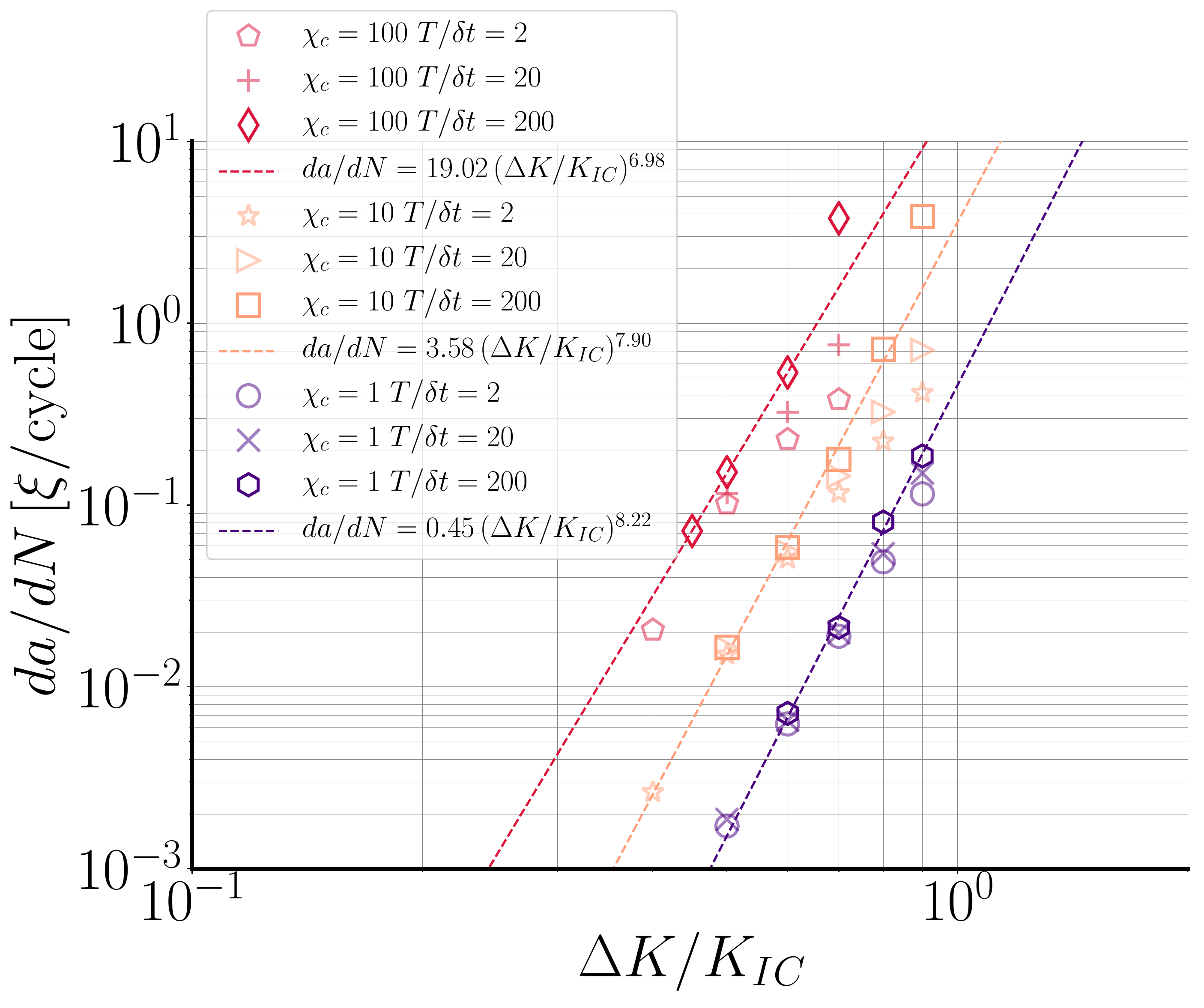}
	\caption{Emergent Paris law in quasi-static simulations of single mode-I crack for the $\gamma$-model~\ref{sub:variable-gamma} showing the convergence for for different number of time steps per cycle.}
	\label{fig:Paris-gamma-qstatic}
\end{figure}

\Figs~\ref{fig:Paris-Ec-qstatic}--\ref{fig:Paris-gamma-qstatic} illustrates the emergent Paris law for different values of $\chi_c$ for the two models. 
We can observe that the exponent of the power-law decreases with increasing $\chi_c$. 
This non-trivial behavior can be interpreted as resulting from the fact that, for sufficiently large $\chi_c$, the degradation variable $\gamma$ reaches its minimum value $\gamma_{\min}$ in a finite region around the crack tip, as found by inspection of the numerical results. The size of this region increases with the amplitude $K_I/K_{IC}$ of cyclic loading. Therefore, as $\chi_c$ increases, the size of the region around the crack tip where $\gamma>\gamma_{\min}$ decreases, thereby decreasing the sensitivity of degradation and crack growth to load $K_I/K_{IC}$.

\Figs~\ref{fig:Paris-Ec-qstatic}--\ref{fig:Paris-gamma-qstatic} show convergence of the rate of crack growth $da/dN$ with increasing number of time steps per cycle ($T/\delta t$ varying from 2 to 200).   
Convergence is faster at lower $\chi_c$ for a given load $\Delta K/K_{IC}$ since a lower degradation rate can be resolved with a smaller number time steps per cycle. As a result, the rate of crack growth is converged for all loads at $\chi_c=1$ but only a smaller range of loads for larger $\chi_c$. We note that, even though resolving the continuum limit of the time-dependent evolution equations \eqref{eq:evolution-ec} and \eqref{eq:evolution-gamma} controlling the degradation process requires the time step to be chosen small enough for a given $\chi_c$ and load, the corresponding discrete evolution equations \eqref{iter1} and \eqref{iter2} still produce Paris law type behavior over a finite range of loads. Hence, similarly as iterated maps of dynamical systems, they provide discrete time models of fatigue crack growth with exponents that decrease with decreasing $\delta t/T$ as seen in \Figs~\ref{fig:Paris-Ec-qstatic}--\ref{fig:Paris-gamma-qstatic}. 


\begin{figure}[htb!]
	\centering
	\includegraphics[width=.9\textwidth]{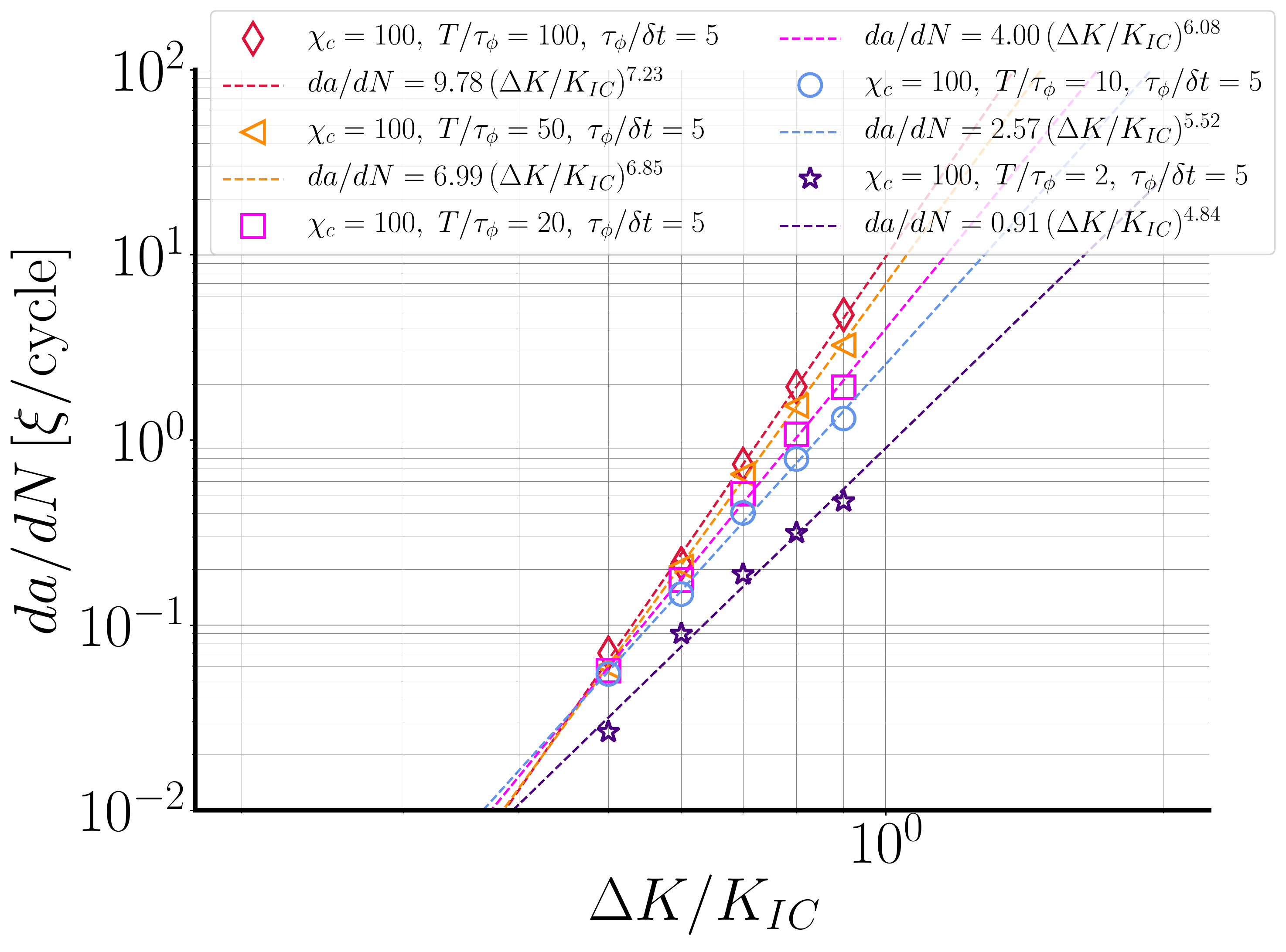}
	\caption{Emergent Paris law in Ginzburg-Landau simulations of single mode-I crack for the variable critical energy model.}
	\label{fig:Paris-Ec-dynamic}
\end{figure}

\begin{figure}[htb!]
	\centering
	\includegraphics[width=.9\textwidth]{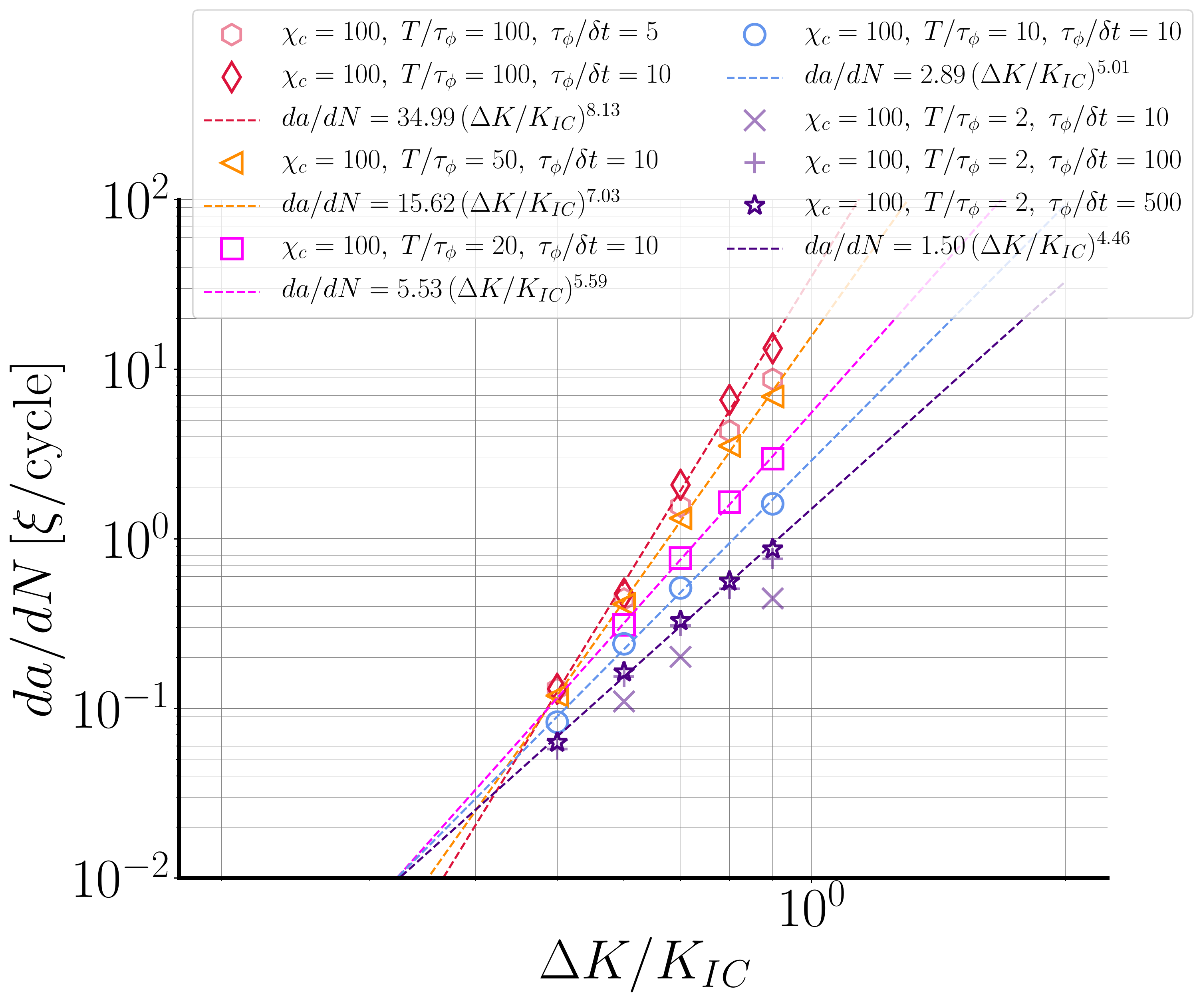}
	\caption{Emergent Paris law in Ginzburg-Landau simulations of single mode-I crack for the $\gamma$-model. The covergence is shown for the two exterme cases $T/\tau_\phi=2,100$ as a function of $\tau_\phi/\delta t$.}
	\label{fig:Paris-gamma-dynamic}
\end{figure}

Using the dynamic formulation~\eqref{eq:gov-phi-gl1} allows for further probing of the models where we set $\delta t=\tau_{\phi}/5$.
As expected, when the relaxation time scale of the phase-field is much smaller than the cyclic loading period ({$\tau_{\phi}\ll T$}), the Paris-law exponent converges to its quasi-static value as depicted in \Fig~\ref{fig:Paris-Ec-dynamic}--\ref{fig:Paris-gamma-dynamic}.
Interestingly, away from this limit (i.e. for $\tau_{\phi}\sim T$), the Paris law exponent is significantly lower than its quasi-static value. This result suggests that dissipation processes that slow down crack growth on the time scale of cyclic loading contribute to decreasing the sensitivity of the crack growth rate to the local fracture toughness degradation rate and hence lowering the Paris law exponent.  

\subsection{Crack interaction}\label{sub:En-passant}

\begin{figure}[htb!]
	\centering
	\includegraphics[height=.3\textheight]{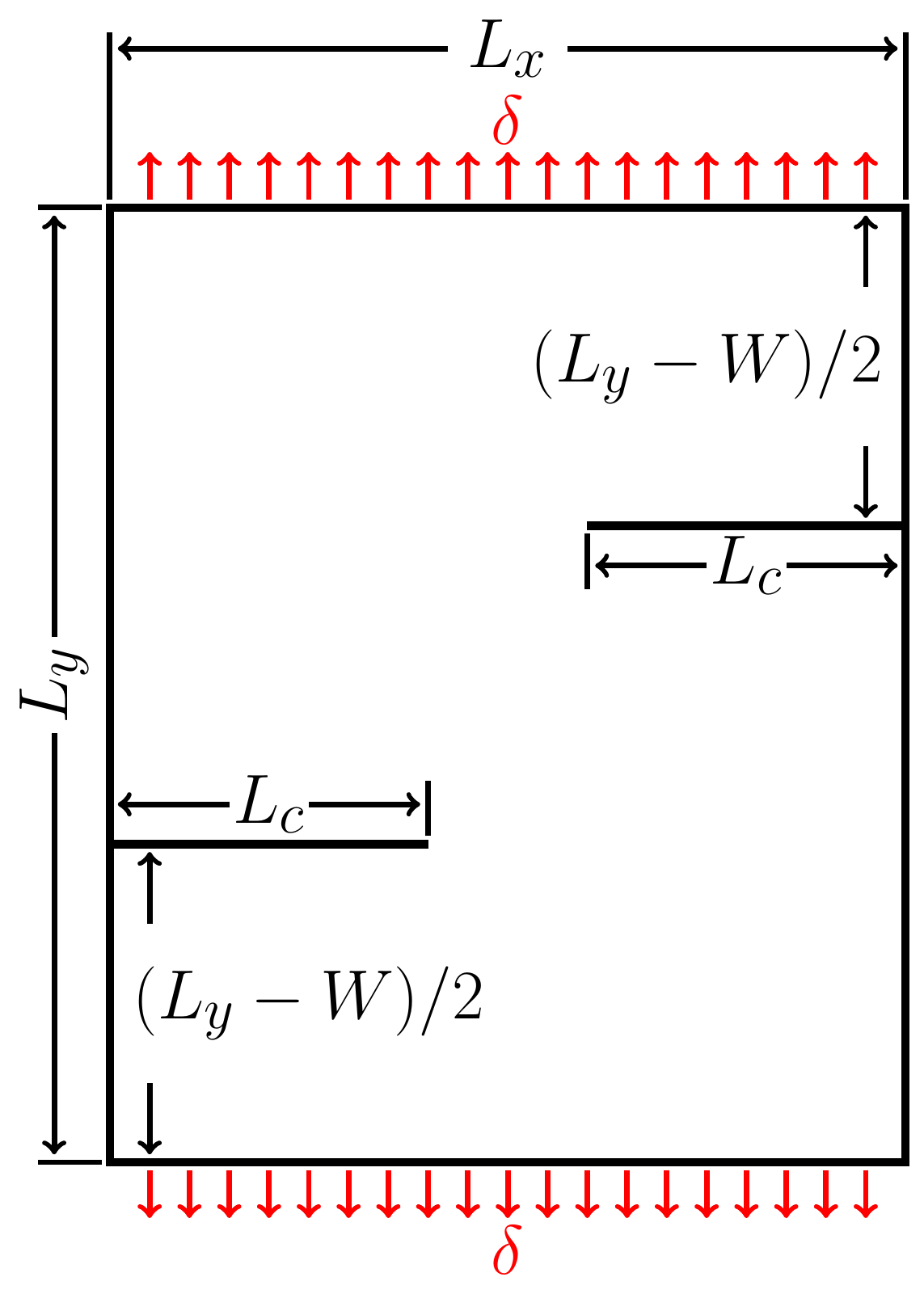}
	\caption{Schematics of the \emph{En-passant} simulations}
	\label{fig:schematics-En-passant}
\end{figure}

In the next set of simulations, we examine the problem of crack interactions for two offset cracks approaching each other under tensile loading. 
The naturally occurring \emph{En-passant} configuration~\cite{Acocella:2000,Peacock:2002,Wesley-Patterson:2010}, has been the focus of a series of theoretical~\cite{Lange:1968,Melin:1983,Ghelichi:2015} and experimental studies~\cite{Fender:2010,Koivisto:2016}.
The geometry of the problem is illustrated in \Fig~\ref{fig:schematics-En-passant} where we apply a vertical displacement $\ub(\pm L_y)=(0,\delta(t))$. Table~\ref{tab:En-passant} summarizes the parameters used for the \emph{En-passant} numerical simulations. 
All simulations were performed using a mesh consisting of $\sim520k$ delaunay triangular elements with an average size of $\xi/5$.

\begin{table}[htb!]
\centering
\begin{tabular}{c|c|c|c|c|c}
$L_y/L_x$ & $L_x/L_c$ & $L_x/W$ & $L_x/\xi$ & $\nu$ & $\delta_{\max}/\sqrt{G_{c}\xi/E}$ \tabularnewline
\hline 
$2$ & $2.5$ & $20$ & $200$  & $0.1$& $14$\tabularnewline
\end{tabular}
\caption{The \emph{En-passant} simulation parameters}
\label{tab:En-passant}
\end{table}

\begin{figure}[htb!]
	\centering
	\includegraphics[width=.83\textwidth]{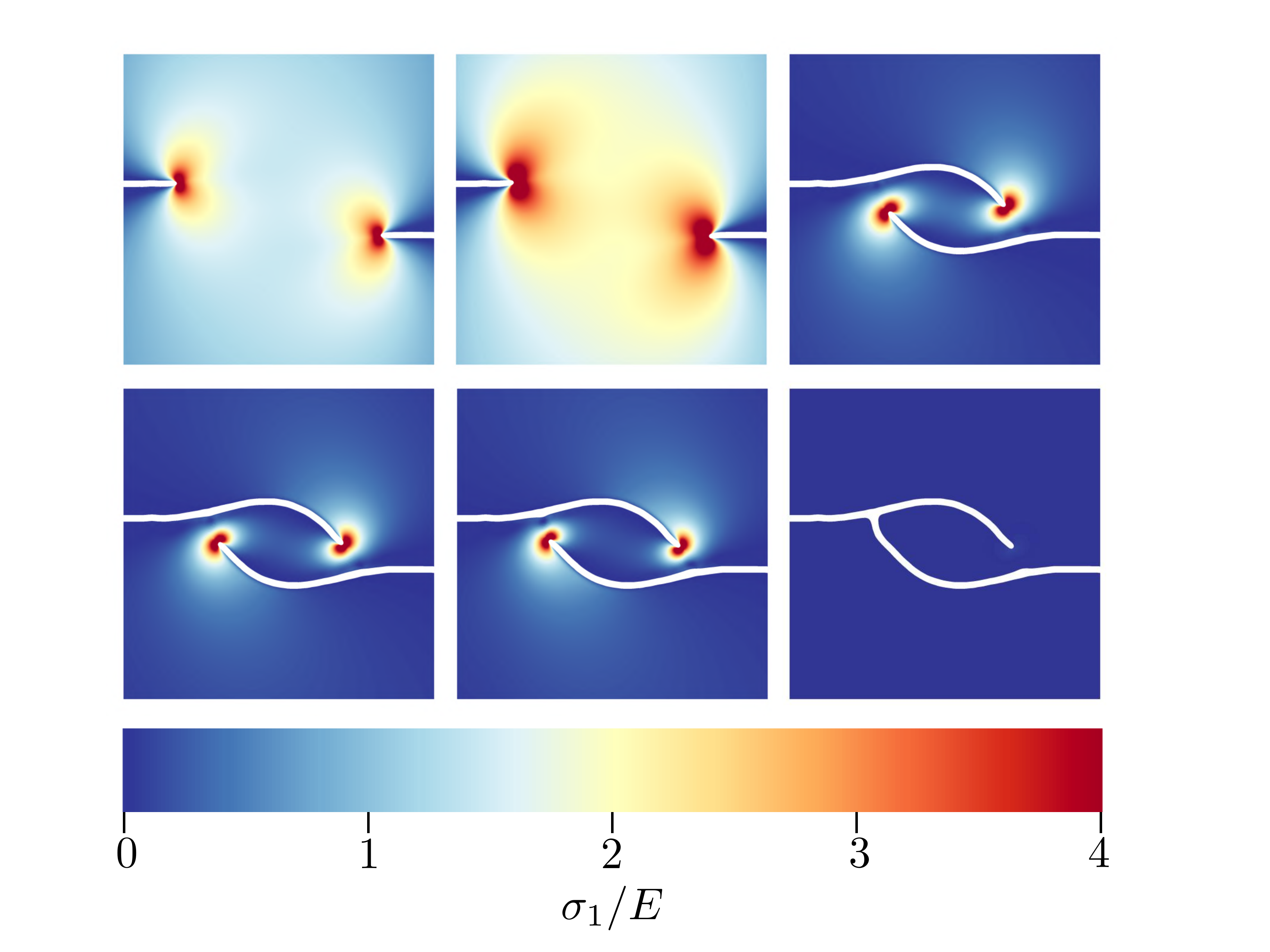}
	\caption{Crack trajectory in quasi-static simulation of \emph{En-passant} experiments $(-0.15\,L_x,0.15\,L_x)^2$ at (from top left) $\delta/\sqrt{G_{c}\xi/E}=15,19.9,20,30,40,42.4$. The color code represents the maximum normal stress $\sigma_1\sqrt{L_x}/K_{IC}$ and the white space represents the crack \ie $\phi\leq0.5$.}
	\label{fig:En-passant-QS-S1}
\end{figure}

It is essential to first perform a quasi-static simulation without degradation to calculate the initial energy release rate $G(\delta)$ (or $K_I$) needed for the fatigue simulations.
\Fig~\ref{fig:En-passant-QS-S1} depicts the results of this \emph{En-passant} quasi-static simulation where the load $\delta$ was increased monotonically ($\delta(t)=kt$).
The setup of the problem is such that each crack first deviates from its original plane and then rotates toward the opposing crack. 
The energy release rate from the crack tip increases at first, destabilizing the crack at $\delta\simeq20\sqrt{G_{c}\xi/E}$.
As the cracks turn, they lose energy (the energy release rate decreases) so that the complete fracture happens at $\delta\simeq42.4\sqrt{G_{c}\xi/E}$.

\begin{figure}[htb!]
	\centering
	\includegraphics[height=.35\textheight]{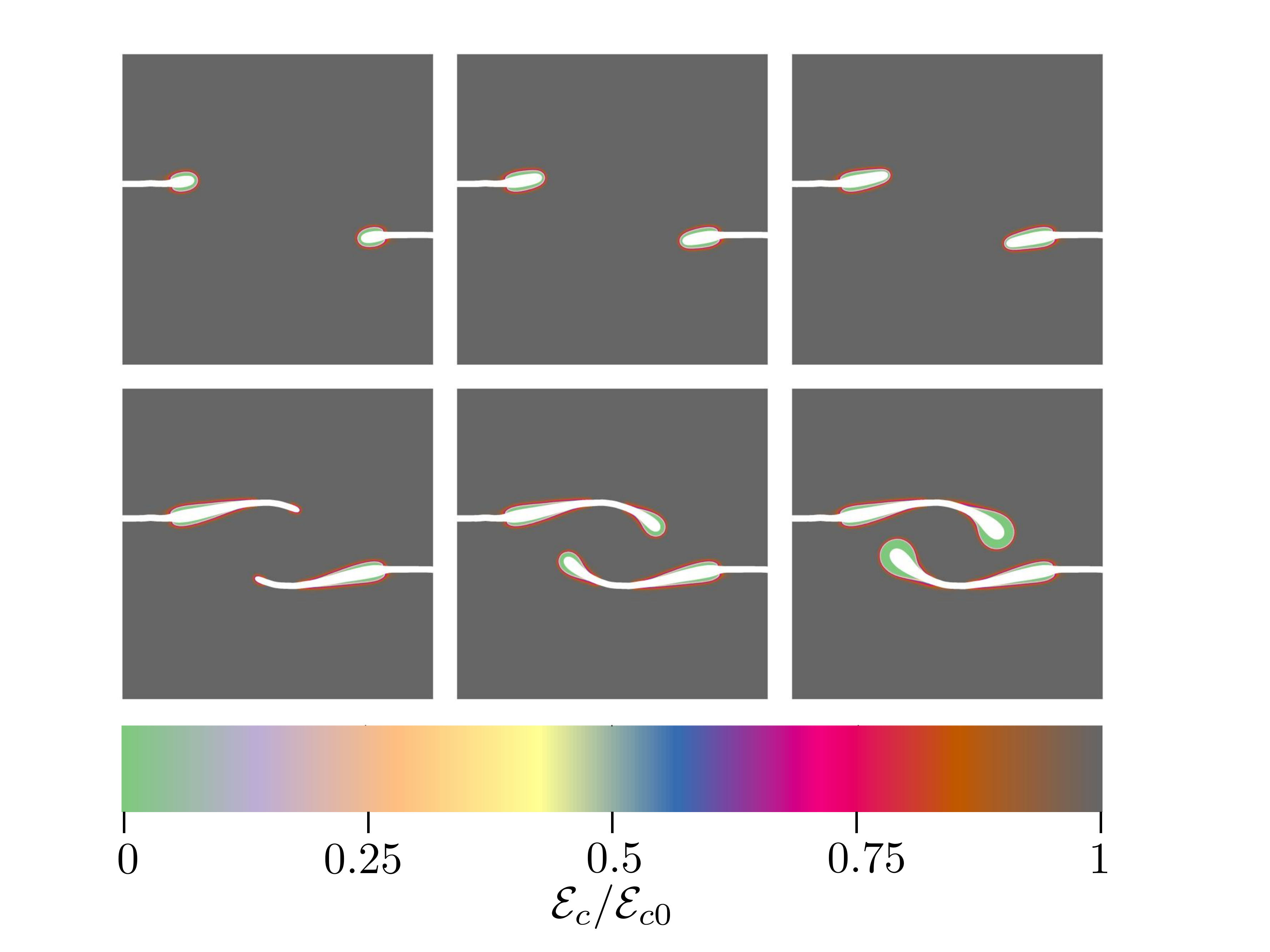}
	\caption{Crack trajectory in simulation of \emph{En-passant} experiments using variable critical energy model at $(-0.15\,L_x,0.15\,L_x)^2$ at cycles (from top left) $250,350,400,450,750,1500$. The color code represents the dimensionless critical energy density $\E_{c}/\E_{c0}$ and the white space represents the crack \ie $\phi\leq0.5$.}
	\label{fig:En-passant-Ec}
\end{figure}

\begin{figure}[htb!]
	\centering
	\includegraphics[height=.35\textheight]{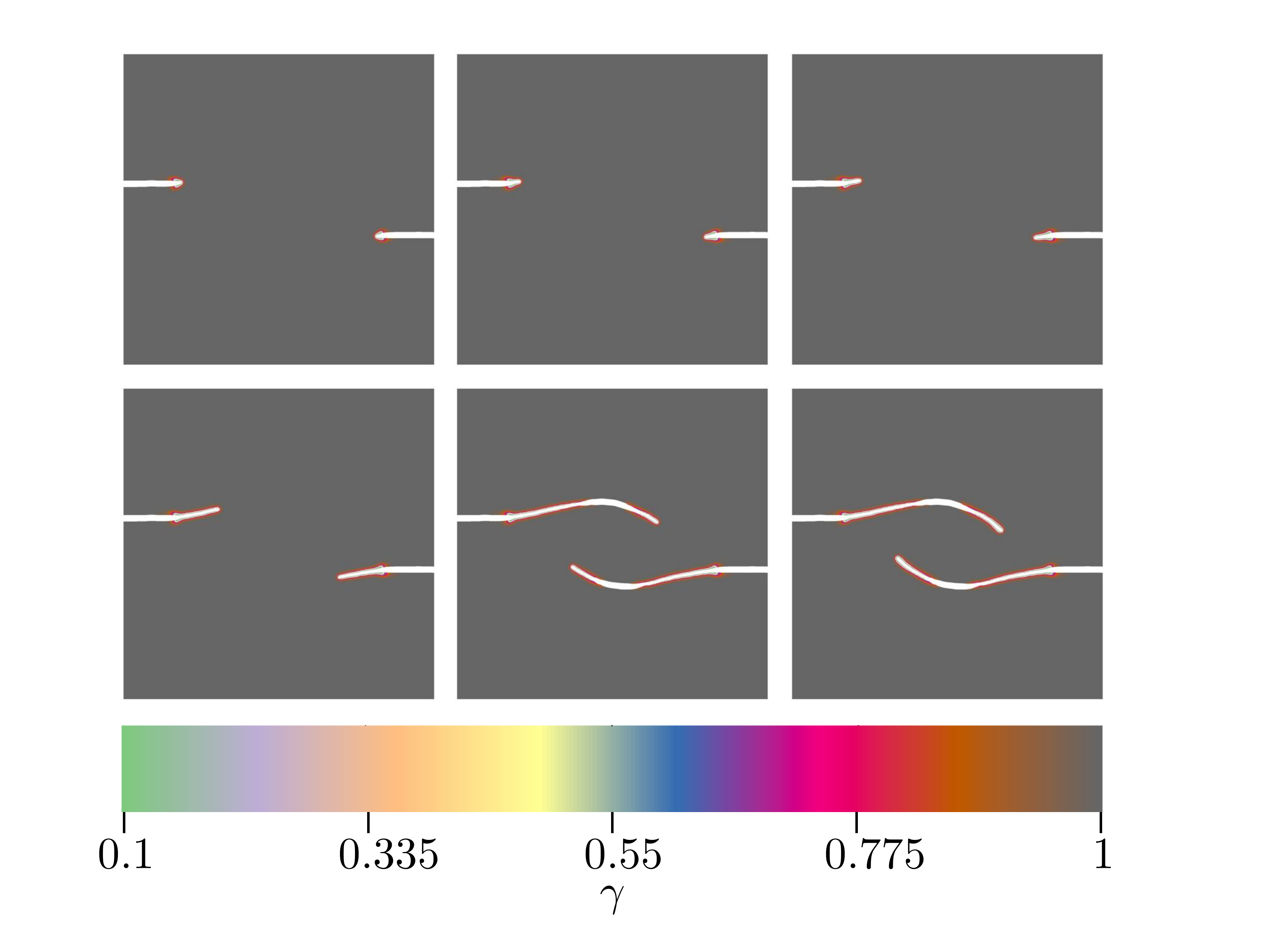}
	\caption{Crack trajectory in simulation of \emph{En-passant} experiments using $\gamma$-model at $(-0.15\,L_x,0.15\,L_x)^2$ at cycles (from top left) $125,250,500,750,1000,1500$. The color code represents the fracture toughness ratio parameter $\gamma$, and the white space represents the crack \ie $\phi\leq0.5$.}
	\label{fig:En-passant-gamma}
\end{figure}

We performed two fatigue simulations corresponding to each model, where $\delta(t)$ oscillates as before in a sawtooth fashion between $0$ and $\delta_{\max}$ in period $T$. Here we set $\delta_{\max}=14\sqrt{G_c \xi/E}$ which corresponds to $K_I/K_{IC}\simeq 0.7$.
We also set the proportionality factors for both simulations to $\chi_c=1$.
Since our models reduce to their quasi-static counterparts at $\E_{c}=\E_{c0}$ or $\gamma=1$, they capture the transition between the fatigue fracture and the quasi-static regime naturally.
Both simulations show that the crack initially propagates at the sub critical load level but then destabilizes once $G(\delta_{\max})\geq G_{c}$ and propagates abruptly (depicted in \Fig~\ref{fig:En-passant-Ec} frames 3-4 and in \Fig~\ref{fig:En-passant-gamma} frames 4-5).
Finally, as the energy release rate from the tip of the turning cracks decreases, the fatigue propagation slows down significantly.


\begin{figure}[htb!]
	\centering
	\includegraphics[width=.65\textwidth]{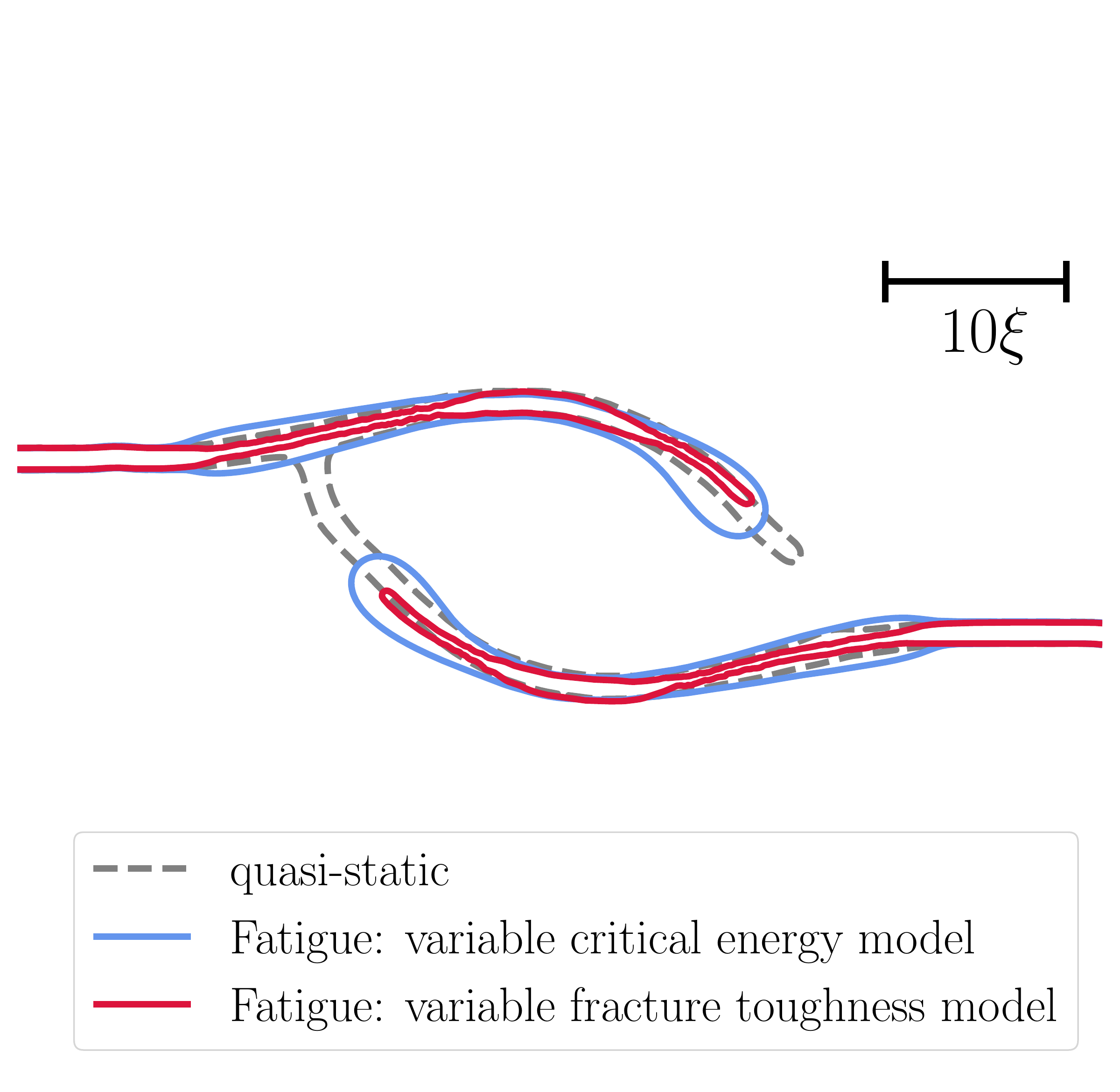}
	\caption{Comparison of crack trajectory (\ie $\phi=0.5$ contour line) in simulation of \emph{En-passant} experiments: quasi-static crack path (gray), variable critical energy model (blue), and $\gamma$-model (red).}
	\label{fig:En-passant-comparison}
\end{figure}

\Fig~\ref{fig:En-passant-comparison} compares the crack path of the quasi-static evolution under monotonic loading with the crack paths obtained from the critical energy and the $\gamma$-model under cyclic loading.
The three macroscopic crack paths are in good agreement up to a spatial resolution comparable to the process zone size. Also, the abrupt propagation segment of both fatigue crack paths remains unchanged compared to the quasi-static fracture path.

\subsection{3-D penny-shaped cracks coalescence}\label{sub:two-penny-shaped-crack}

In the first three-dimensional set of simulations, we model the coalescence of two 3-D penny-shaped cracks. Two penny-shaped cracks with radius $R_c$ are initialized at distance $b$ in an $L^3$ cube where we apply displacement boundary conditions $\ub=(0,0,\delta(t))$ at $(x,y,\pm L)$. 
A similar problem was originally investigated  by Legrand~\emph{et al.}~\cite{Legrand:2015} using a semi-analytic method where the authors used a perturbation method along with the principle of local symmetry (i.e. vanishing mode II stress intensity factor at the crack tip) to predict the propagation and coalescence of the cracks.
We performed this simulation using a mesh containing $16.5\,M$~tetrahedral elements and the $\gamma$-model~\ref{sub:variable-gamma}.
Since the computational cost of performing a quasi-static simulation (over many cycles) is prohibitive, we opted for the dynamic phase-field formulation~\ref{sub:ginzburg-landau} with the loading period $T/\tau_{\phi}=10$ and $\chi_{c}=10$. 
The simulation parameters are summarized in table~\ref{tab:penny-shaped-crack}. 
In this geometry, the energy release rate increases as the cracks grow and speeds up the fatigue crack propagation. 
Furthermore, the energy release rate increase is more dramatic at the coalescence site resulting in the cracks merging over 5 cycles between $N=50-55$. 
This is similar to the behavior predicted by Legrand~\emph{et al.} (see for example Fig.~5 in~\cite{Legrand:2015}).

\begin{figure}[htb!]
	\centering
	\includegraphics[width=.5\textwidth]{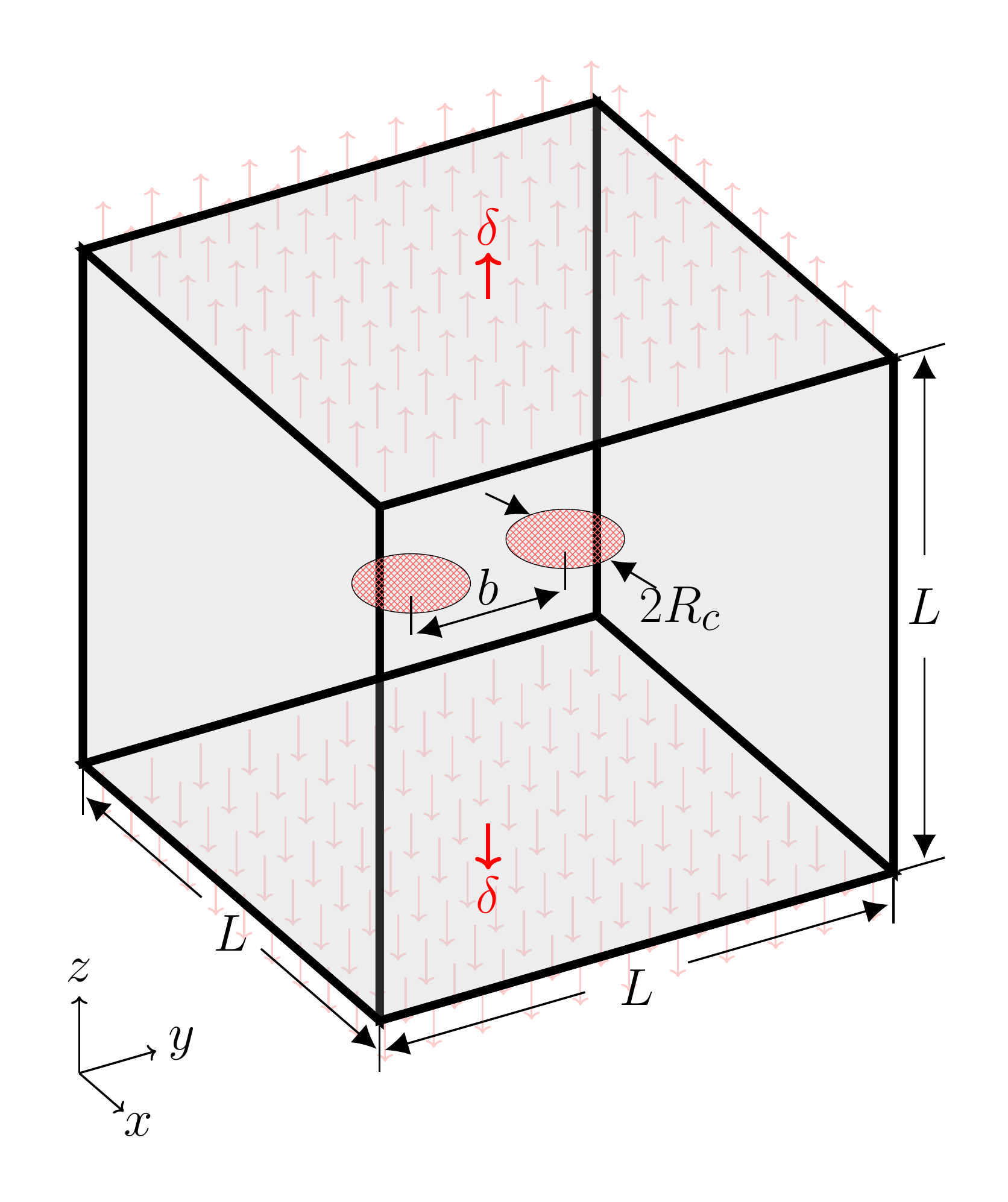}
	\caption{Schematics for the 3-D penny-shaped cracks coalescence problem.}
	\label{fig:schematics-2penny-shaped-crack}
\end{figure} 

\begin{table}[htb!]
\centering
\begin{tabular}{c|c|c|c|c|c}
$L/\xi$ & $R_c/\xi$ & $b/\xi$ & $\nu$ & $\gamma_{\min}$ &  $\delta_{\max}/\sqrt{G_{c}\xi/E}$\tabularnewline
\hline
$50$ & $5$  & $20$ & $1/3$ & $0.5$ & $6.36$ \tabularnewline
\end{tabular}
\caption{Simulation parameters for 3-D penny-shaped cracks coalescence}
\label{tab:2penny-shaped-crack}
\end{table}

\begin{figure}[htb!]
	\centering
	\includegraphics[width=.9\textwidth]{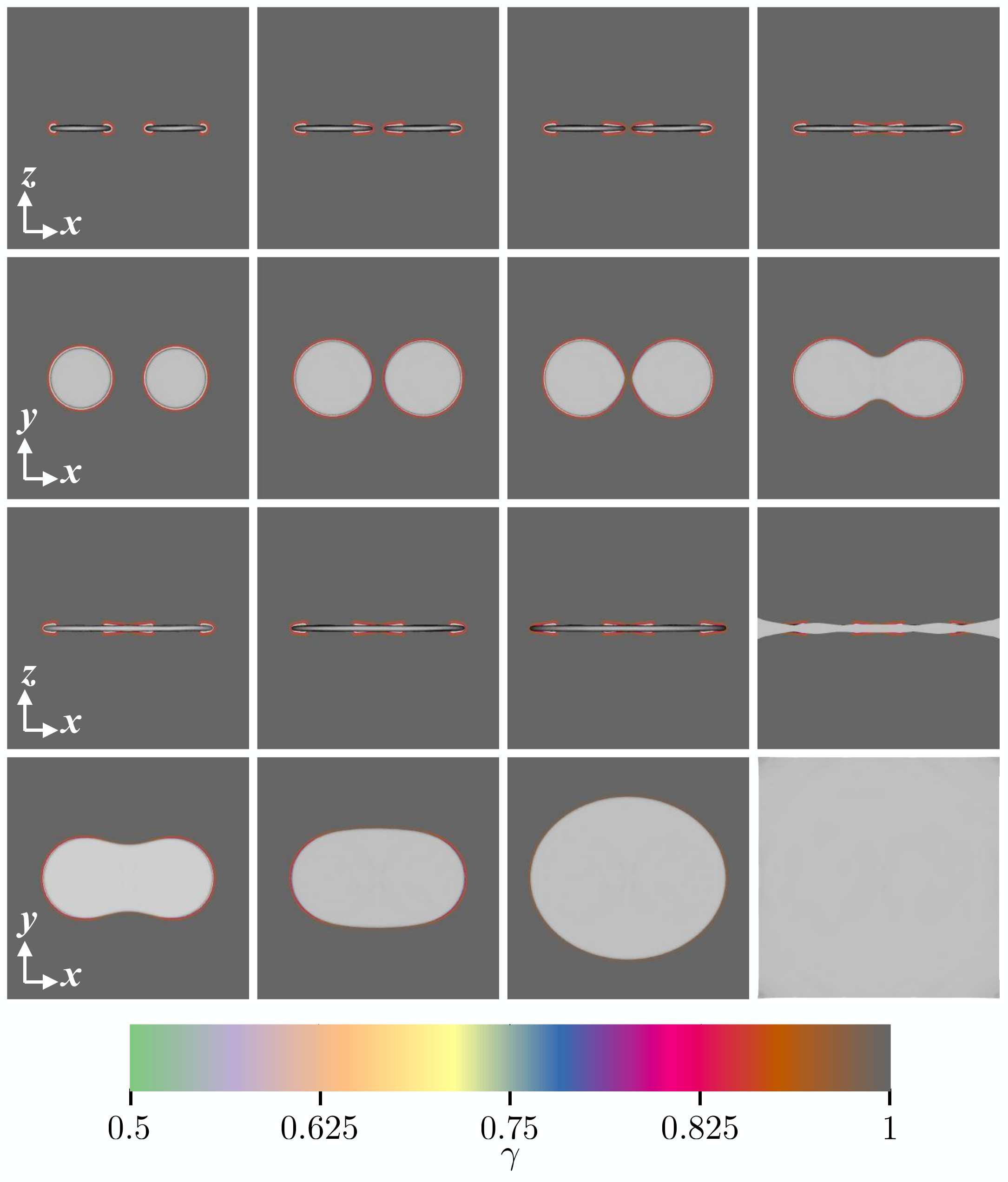}
	\caption{Results of the 3-D penny-shaped cracks coalescence simulation at cycles (from top left) $45,50,51,52,53,55,60,100$ (first and third row) side view (second and fourth row) top view. The color code represents the fracture toughness ratio parameter $\gamma$ (a small amount of transparency was introduced to highlight 3-D fracture topology), and the gray volume represents the crack \ie $\phi\leq0.5$.}
	\label{fig:2penny-shaped-crack-gamma}
\end{figure} 

\subsection{3-D inclined penny-shaped crack}\label{sub:penny-shaped-crack}

In the last set of numerical simulations, we model the propagation of a 3-D penny shaped crack. A tilted penny-shaped crack with radius $R_c$ is initialized in the middle of an $L^3$ cube with its normal $\boldsymbol{n}_c=\left(\cos(\theta_c),\cos(\theta_c),\sin(\theta_c)\right)$. The loading $\ub=(0,0,\delta(t))$ is applied at $(x,y,\pm L)$. 
A similar problem was originally simulated by Gravouil~\emph{et al.}~\cite{Gravouil:2002} using a level set method in a quasi-static setting for $\theta=\pi/4$.
We performed this simulation using a mesh containing $16\,M$~tetrahedral elements. The simulation was performed using the $\gamma$-model~\ref{sub:variable-gamma} and the dynamic phase-field formulation~\ref{sub:ginzburg-landau} (similar to the previous section to limit the computational cost).
We chose the loading period $T=10\,\tau_{\phi}$ and $\chi_{c}=10$. The simulation parameters are summarized in table~\ref{tab:penny-shaped-crack}.

\begin{figure}[htb!]
	\centering
	\includegraphics[width=.5\textwidth]{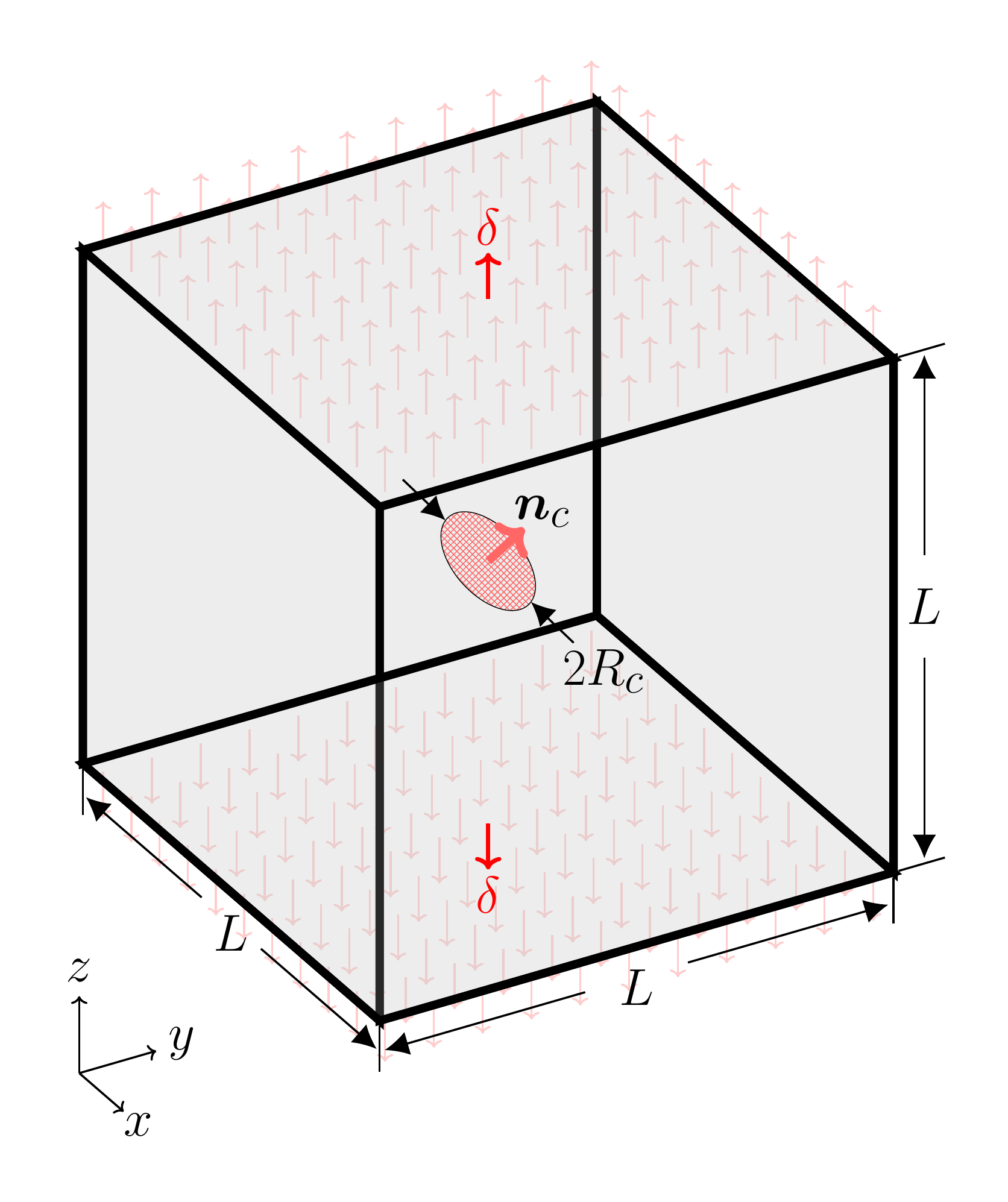}
	\caption{Schematics of 3-D penny-shaped crack geometry.}
	\label{fig:schematics-penny-shaped-crack}
\end{figure} 

\begin{table}[htb!]
\centering
\begin{tabular}{c|c|c|c|c|c}
$L/\xi$ & $R_c/\xi$ & $\nu$ & $\theta$ & $\gamma_{\min}$ & $\delta_{\max}/\sqrt{G_{c}\xi/E}$\tabularnewline
\hline
$50$ & $5$  & $1/3$& $\pi/3$&$0.1$&$5.66$\tabularnewline
\end{tabular}
\caption{Simulation parameters for the 3-D inclined penny-shaped crack}
\label{tab:penny-shaped-crack}
\end{table}

\begin{figure}[htb!]
	\centering
	\includegraphics[width=.9\textwidth]{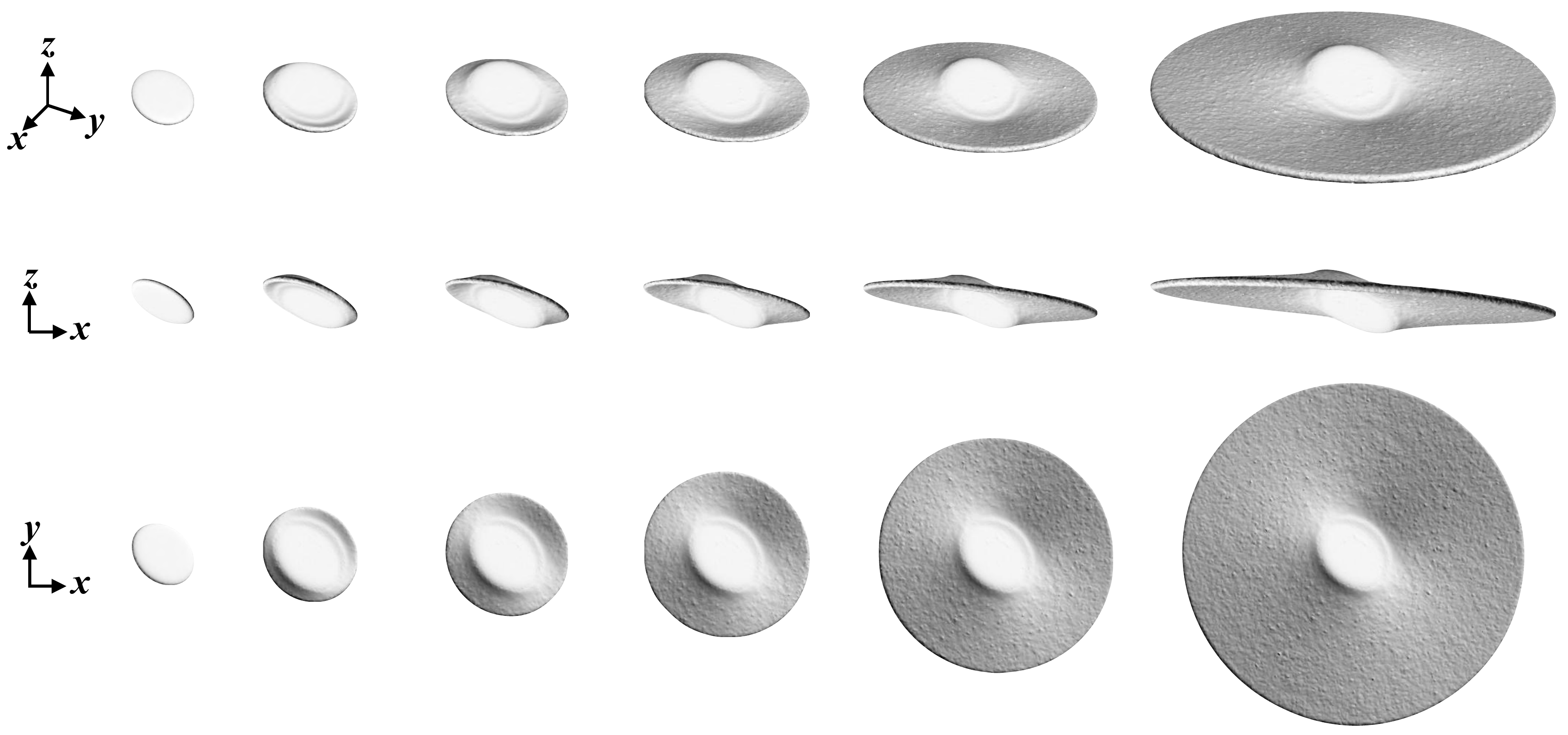}
	\caption{Crack path (\ie $\phi\leq0.5$) of the 3-D penny-shaped crack simulation for $\theta_c=\pi/3$ at cycles (from left) $0,20,40,60,80,100$ (top) isometric view (middle) side view (bottom) top view.}
	\label{fig:penny-shaped-crack-alpha}
\end{figure}

\begin{figure}[htb!]
	\centering
	\includegraphics[width=.9\textwidth]{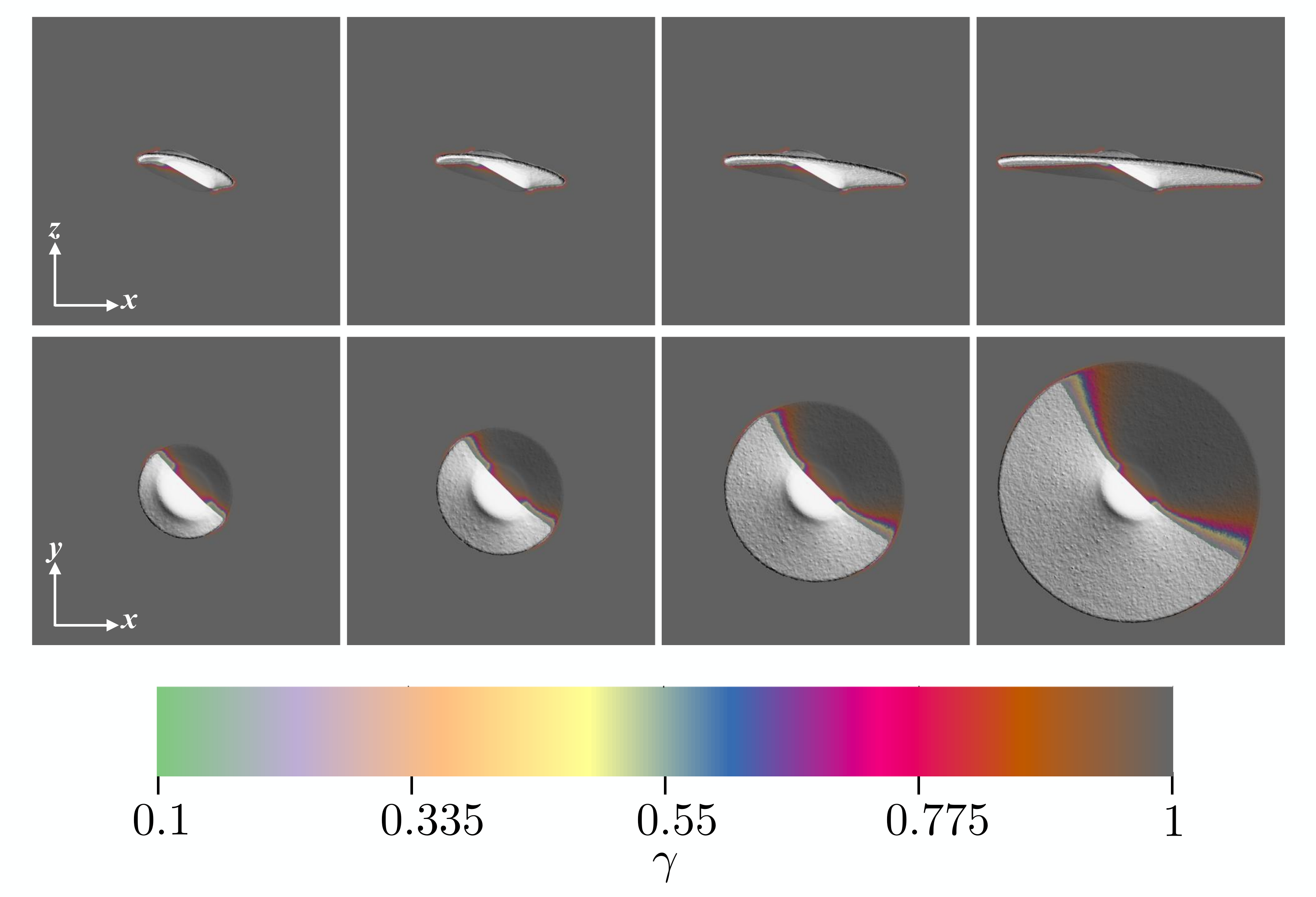}
	\caption{Results of the 3-D penny-shaped crack simulation for $\theta_c=\pi/3$ at cycles (from left) $40,60,80,100$ (top) side view (bottom) top view. The color code represents the fracture toughness ratio parameter $\gamma$ (a small amount of transparency was introduced to highlight 3-D fracture topology), and the white volume represents the crack \ie $\phi\leq0.5$.}
	\label{fig:penny-shaped-crack-gamma}
\end{figure}

\Fig~\ref{fig:penny-shaped-crack-alpha} shows the crack front obtained where, similar to brittle fracture, the crack front twists and turns towards the symmetry line $z=0$ where its $K_{II}=0$. 
As shown in this figure, the propagation of the fatigue crack accelerates over time. 
This is due to the increase of the energy release rate from the crack tip, which can be easily understood since the area under tension decreases as the crack propagates.
\Fig~\ref{fig:penny-shaped-crack-gamma} shows how the relative fracture toughness ratio $\gamma$ changes spatially around the crack front. 
Like the \emph{En-passant} simulations in section~\ref{tab:En-passant}, the change in $\gamma$ is limited to the immediate vicinity of the crack front. 

\begin{figure}[htb!]
	\centering
	\includegraphics[width=.5\textwidth]{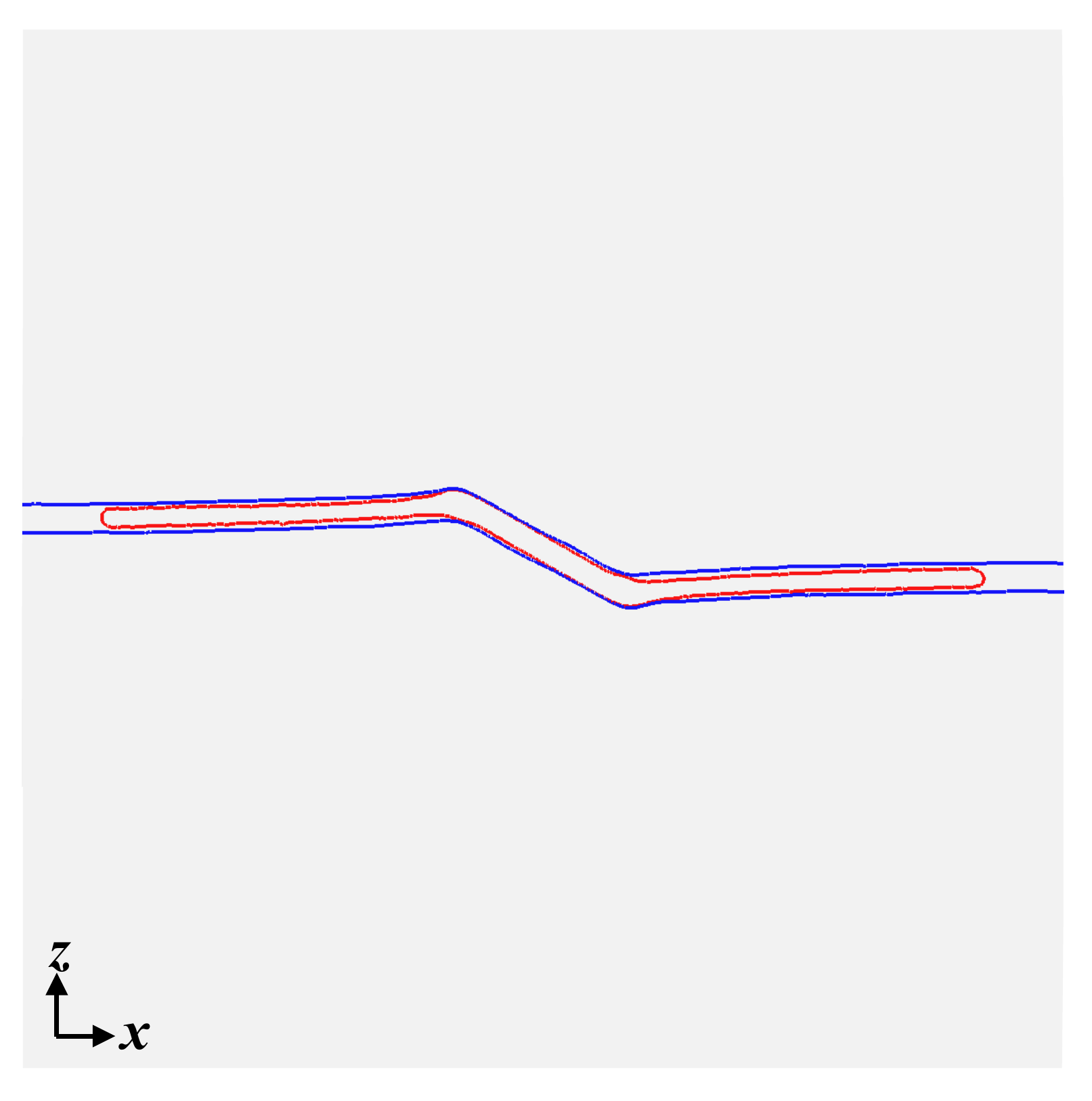}
	\caption{Results of the 3-D penny-shaped crack simulation for $\theta_c=\pi/3$: comparison of the crack path (\ie $\phi=0.5$ contour line) on X-Z plane for quasi-static simulation (blue) at $\delta/\sqrt{G_{c}\xi/E}=14.14$ vs. the $\gamma$-model (red) at the end of $100^{th}$ loading cycle.}
	\label{fig:penny-shaped-crack-comparison}
\end{figure}

Finally, \Fig~\ref{fig:penny-shaped-crack-comparison} shows a comparison between the quasi-static and fatigue crack paths. As in the case of the \emph{En-passant} simulations~\ref{sub:En-passant}, the crack path remains unaltered with respect to the quasi-static path.

\section{Towards a ductile phase-field fatigue formulation}\label{sec:non-local}

{In all previous sections, the fatigue degradation was limited to the process zone around the crack tip. 
However, in \emph{K-dominant} fatigue of ductile materials, the crack-tip plastic zone has a scale larger than the fracture process zone but much smaller than the specimen dimensions.
Furthermore, ductile fracture is driven by void nucleation, growth, and coalescence mediated by the plasticity. 
Locally, voids reduce the load-bearing portion of the crack plane similar to a damage variable in the continuum description. 
Indeed, such modifications have been suggested and implemented in the celebrated Gurson-Tvergaard-Needleman (\emph{GTN}) \cite{Gurson:1977a,Needleman:1984} model to account for the effect of plasticity on fracture \cite{Xue:2008}. 
Thus, one can posit that in ductile fracture, the fracture toughness locally reduces due to plasticity. Such a ductile phase-field fracture formulation has been proposed in \cite{Ambati:2015} where $g(\phi)$ was modified in~\eqref{eq:ATE-Bulk} to include the equivalent plastic strain in a way that lowers the fracture toughness locally in plastically deforming areas and retards the creation of cracks in the elastic regions.}

Here we suggest a simple extension of the previously developed fracture toughness model to account for degradation on a scale larger than the fracture process zone. We modify the evolution equation for $\gamma$~\eqref{eq:evolution-gamma} to allow for fatigue degradation of fracture toughness to take place at any point where the tensile elastic energy potential is greater than a critical stress $W^{+}>\sigma_c$ \ie
\begin{equation}\label{eq:evolution-gamma-nonlocal}
	\frac{1}{\chi_{c}}\frac{\partial \gamma}{\partial t}  =\min\left(0, -\mathrm{H}(\gamma-\gamma_{\min})\mathrm{H}(W^{+}-\sigma_c) \frac{\partial}{\partial t} \left(g(\phi)\frac{W^+}{\E_{c0}} \right)\right)
\end{equation}

\Fig~\ref{fig:Paris-nonlocal} shows the results of the simulations for a simple mode-I geometry depicted earlier in figure~\ref{fig:schematics-favorite-geom}. 
For all simulations shown, we set $\chi_c=1$, $\nu=1/3$ and $L_y/L_x=1/4$. 
As one can see, the Paris law is reproduced with a smaller exponent for smaller $\sigma_c$.
Comparing $\sigma_c$ to the divergent stresses at the crack tip, one can derive the fatigue degradation zone's radius as
\begin{equation}
	r_c=\frac{1}{2\pi}\left(\frac{K_{IC}}{\sigma_c}\right)^2
\end{equation}
\Fig~\ref{fig:Paris-nonlocal} shows that the modified degradation law \eqref{eq:evolution-gamma-nonlocal} reproduces the Paris law with an exponent that decreases with increasing radius $r_c/\xi$ of the degradation zone. Importantly, when $r_c\gg \xi$, this exponent ($m\approx 4.64$ when $r_c/\xi=32$) is substantially smaller than the exponent ($m\approx 8.22$) obtained when degradation is localized to the process zone (\Fig~\ref{fig:Paris-gamma-qstatic}). As expected, when $r_c$ is comparable to $\xi$, the exponent is similar to the one previously obtained. 

By extending this approach, we have developed a ductile fatigue model where the degradation of the fracture toughness is controlled by the accumulated plastic strain $e^p_{eq}$, which is achieved by replacing \eqref{eq:evolution-gamma-nonlocal} by
\begin{equation}\label{eq:gamma-flow-rule-plastic}
	\gamma=\max(\gamma_{\min},1-\chi_{c}\,e^p_{eq})	
\end{equation}
This model reproduces non-trivial features of ductile fatigue such as crack closure. Both the model and results will be presented elsewhere.

\begin{figure}[htb!]
	\centering
	\includegraphics[width=.95\textwidth]{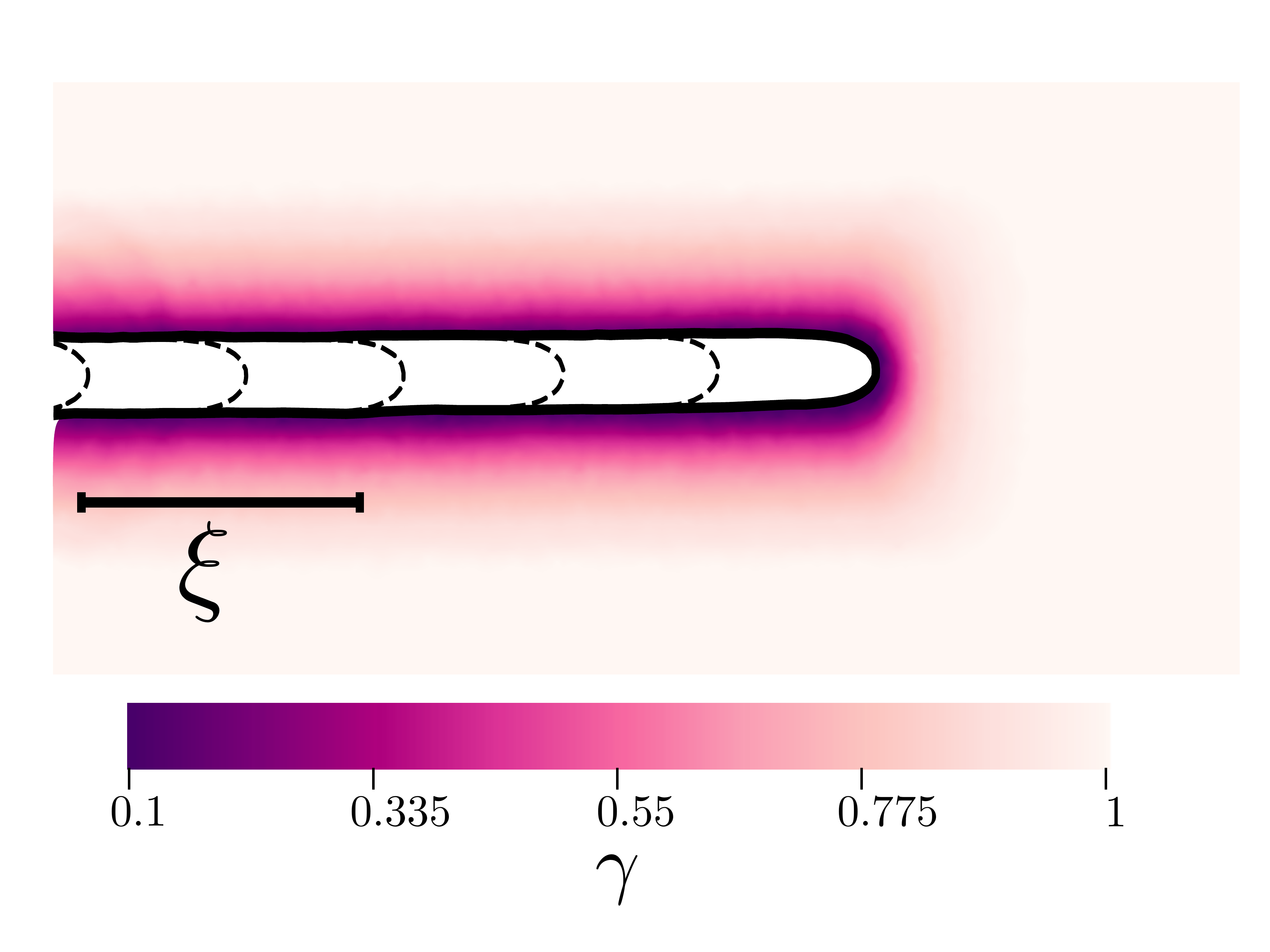}
	\caption{Crack tip (\ie contours of $\phi=0.5$) propagation under mode-I cyclic loading (see figure~\ref{fig:schematics-favorite-geom}) for non-local degradation zone model~\ref{sec:non-local} using $r_c/\xi=0.42 $ for $L_x/L_y=4$, $\nu=1/3$, $\chi_{c}=1$, and $K_I/K_{IC}=3/4$  plotted for 25 cycles at 5 cycle intervals. The color code depicts the fracture toughness ratio field $\gamma$ for the last cycle outside the crack (\ie $\phi\geq0.5$).}
	\label{fig:nonlocal-snapshot}
\end{figure} 

\begin{figure}[htb!]
	\centering
	\includegraphics[width=.95\textwidth]{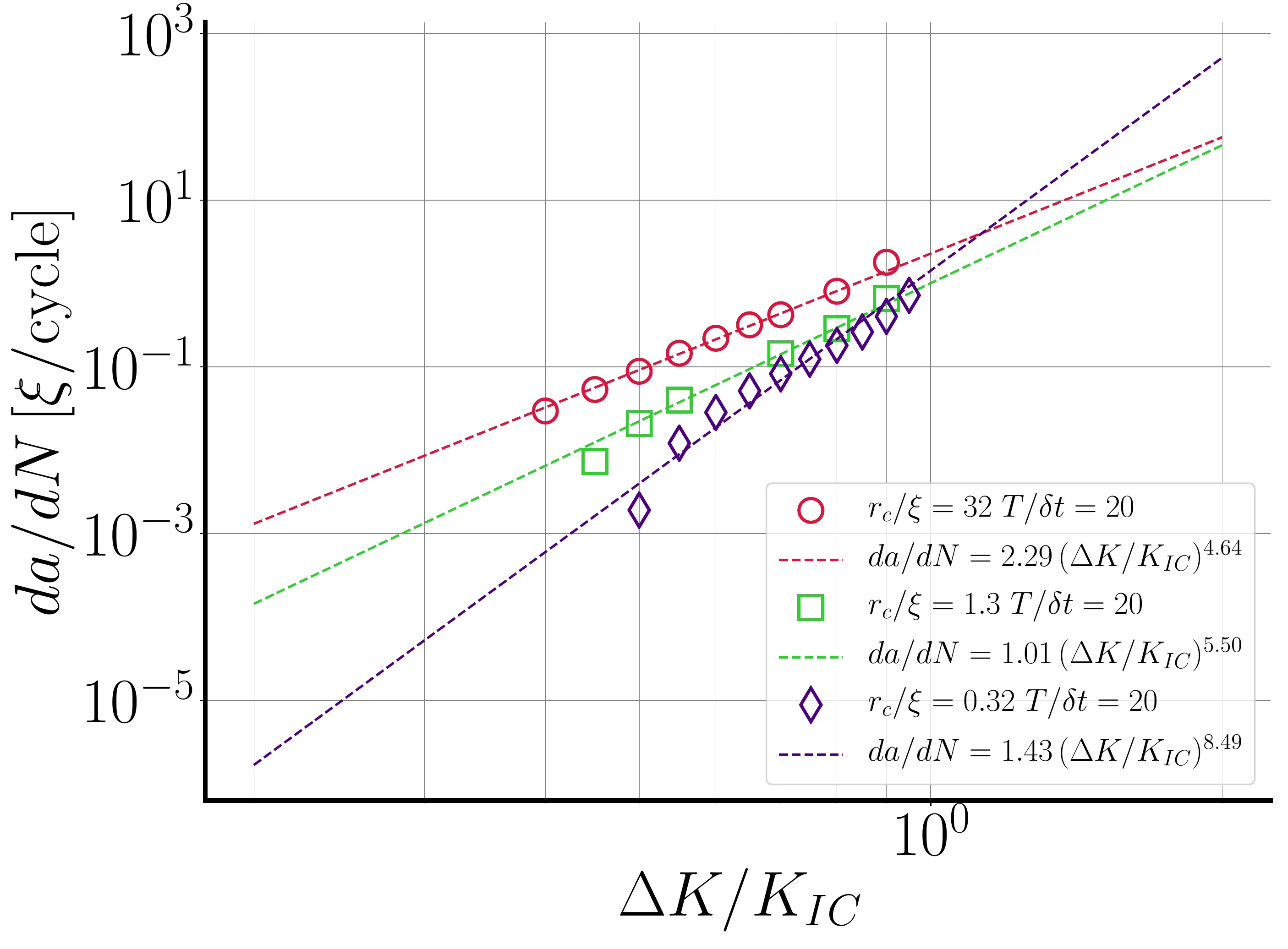}
	\caption{Emergent Paris law in quasi-static simulations of single mode-I crack for critical stress model with non-local degradation zone model~\ref{sec:non-local} using $\chi_c=1$.}
	\label{fig:Paris-nonlocal}
\end{figure} 

\section{Conclusions}\label{sec:conclusions}

We introduced two phase-field models of fatigue crack growth. In our models, the material degrades near the crack tip as a result of cyclic loading which allows for sub-critical (\ie $G<G_{c}$) propagation of cracks.
Crucially, this class of models preserves the fracture toughness as a material property globally and only affects the near-tip properties on the process zone scale. 
In other words, if the crack is loaded beyond $G_{c}$($K_{IC0}$), it propagates at a constant (non-cyclical) load.
This allows us to perform uniform simulations which seamlessly switch between monotonic $G>G_{c}$ and fatigue $G<G_{c}$ crack propagations (see, for example, figure~\ref{fig:En-passant-comparison} in section~\ref{sub:En-passant}).

The most intriguing finding of this article is the emergence of the Paris law. 
We showed (using numerical simulations) that the Paris power-law emerges naturally from our formulation under simple mode-I loading section~\ref{sub:paris-law} in both quasi-static simulations and in simulations where the phase-field $\phi$ follows Ginzburg-Landau dynamics.
In the quasi-static simulations, we recovered the Paris law with high exponents typical of fatigue crack growth in brittle materials~\cite{Ritchie:1999}.
Additionally, we were able to capture the Paris law over a wider range of exponents in Ginzburg-Landau gradient dynamics by changing the ratio of the cyclic loading period over the charachteristic relaxation time of the phase-field $T/\tau_\phi$.

Here, we should also emphasize that our two models should be thought of as two specific cases of a more general class of fatigue phase-field models where the fracture toughness is degraded locally around the crack tip as a function of the elastic power (the time derivative of elastic energy density) stored at each point in the system. 
For example, our choice of linear degradation of $\E_{c}$ and $\gamma$ are the simplest cases, and replacing $H(\E_{c}-\E_{c\min})$ with $(\E_{c}-\E_{c\min})$ in \eqref{eq:evolution-ec} (or, similarly, replacing  $H(\gamma_c-\gamma_{\min})$ with $(\gamma-\gamma_{\min})$ in~\eqref{eq:evolution-gamma}) produces similar results to those presented in this article.

The \emph{En-passant} simulations \ref{sub:En-passant} and the inclined penny-shaped 3-D crack simulations section~\ref{sub:penny-shaped-crack} suggest that, in an isotropic elastic solid, the fatigue crack propagates on a path similar to quasi-static fracture \ie at $K_{II}=0$~\cite{Hakim:2009}.
Even though this observation agrees with current wisdom, it is not formally derived here from an analysis of the phase-field models, as previously done for quasi-static propagation under constant loading \cite{Hakim:2009}. 

Finally, in section~\ref{sec:non-local}, using a simple extension of our brittle model, we showed that in a ductile material where the degradation occurs at the crack-tip plastic zone (on a scale larger than the process zone $\xi$, but smaller than the system size) the Paris law exponent is reduced ($m\simeq4.64$ in~\ref{fig:Paris-nonlocal}).
This key finding paves the way for the introduction of ductile fatigue phase-field models where degradation is driven by plastic deformation.

\section{Acknowledgments}
The authors' work was supported by the U.S. Navy's SBIR/STTR office, under the auspices of contract N68335-16-C-0206. The majority of the numerical simulations were performed using resources of Northeastern University's Discovery cluster located in Massachusetts Green High Performance Computing Center (MGHPCC) in Holyoke, MA.

\pagebreak

\begin{thebibliography}{10}

\bibitem{Abdelmoula:2010}
Radhi Abdelmoula, Jean-Jacques Marigo, and Thibaut Weller.
\newblock Construction and justification of paris-like fatigue laws from
  dugdale-type cohesive models.
\newblock {\em Annals of Solid and Structural Mechanics}, 1(3-4):139--158,
  2010.

\bibitem{Acocella:2000}
Valerio Acocella, Agust Gudmundsson, and Renato Funiciello.
\newblock Interaction and linkage of extension fractures and normal faults:
  examples from the rift zone of iceland.
\newblock {\em Journal of Structural Geology}, 22(9):1233--1246, 2000.

\bibitem{Alessi:2018a}
Roberto Alessi, Stefano Vidoli, and Laura De~Lorenzis.
\newblock A phenomenological approach to fatigue with a variational phase-field
  model: The one-dimensional case.
\newblock {\em Engineering Fracture Mechanics}, 190:53--73, 2018.

\bibitem{Alsem:2007}
Daan~Hein Alsem, Olivier~N Pierron, Eric~A Stach, Christopher~L Muhlstein, and
  Robert~O Ritchie.
\newblock Mechanisms for fatigue of micron-scale silicon structural films.
\newblock {\em Advanced Engineering Materials}, 9(1-2):15--30, 2007.

\bibitem{Alsem:2008}
DH~Alsem, CL~Muhlstein, EA~Stach, and RO~Ritchie.
\newblock Further considerations on the high-cycle fatigue of micron-scale
  polycrystalline silicon.
\newblock {\em Scripta Materialia}, 59(9):931--935, 2008.

\bibitem{Ambati:2015}
M.~Ambati, T.~Gerasimov, and L.~De~Lorenzis.
\newblock Phase-field modeling of ductile fracture.
\newblock pages 1--24, 2015.

\bibitem{Amor:2009}
H.~Amor, J.-J. Marigo, and C.~Maurini.
\newblock {Regularized formulation of the variational brittle fracture with
  unilateral contact: Numerical experiments}.
\newblock {\em Journal of the Mechanics and Physics of Solids},
  57(8):1209--1229, 2009.

\bibitem{Amor:2015}
Hanen Amor, Jean-Jacques Marigo, and Corrado Maurini.
\newblock Numerical experiments in a variational formulation of fatigue.
\newblock 2015.

\bibitem{petsc-user-ref}
Satish Balay, Shrirang Abhyankar, Mark~F. Adams, Jed Brown, Peter Brune, Kris
  Buschelman, Lisandro Dalcin, Victor Eijkhout, William~D. Gropp, Dinesh
  Kaushik, Matthew~G. Knepley, Dave~A. May, Lois~Curfman McInnes, Richard~Tran
  Mills, Todd Munson, Karl Rupp, Patrick Sanan, Barry~F. Smith, Stefano
  Zampini, Hong Zhang, and Hong Zhang.
\newblock {PETS}c users manual.
\newblock Technical Report ANL-95/11 - Revision 3.9, Argonne National
  Laboratory, 2018.

\bibitem{petsc-efficient}
Satish Balay, William~D. Gropp, Lois~Curfman McInnes, and Barry~F. Smith.
\newblock Efficient management of parallelism in object oriented numerical
  software libraries.
\newblock In E.~Arge, A.~M. Bruaset, and H.~P. Langtangen, editors, {\em Modern
  Software Tools in Scientific Computing}, pages 163--202. Birkh{\"{a}}user
  Press, 1997.

\bibitem{Barenblatt:2006}
G~Barenblatt.
\newblock Scaling phenomena in fatigue and fracture.
\newblock {\em Advances in Fracture Research}, pages 19--35, 2006.

\bibitem{Bazant:2014}
Zdenek~P. Bazant and Mija~H. Hubler.
\newblock Theory of cyclic creep of concrete based on paris law for fatigue
  growth of subcritical microcracks.
\newblock {\em Journal of the Mechanics and Physics of Solids}, 63:187--200,
  2014.

\bibitem{Bazant:1991}
Zdenek~P Bazant and Kangming Xu.
\newblock Size effect in fatigue fracture of concrete.
\newblock {\em ACI Mater. J}, 88(4):390--399, 1991.

\bibitem{Benson:2006}
S.~J. Benson and T.~S. Munson.
\newblock Flexible complementarity solvers for large-scale applications.
\newblock {\em Optimization Methods and Software}, 21(1):155--168, 2006.

\bibitem{Borden:2016}
Michael~J Borden, Thomas~JR Hughes, Chad~M Landis, Amin Anvari, and Isaac~J
  Lee.
\newblock A phase-field formulation for fracture in ductile materials: Finite
  deformation balance law derivation, plastic degradation, and stress
  triaxiality effects.
\newblock {\em Computer Methods in Applied Mechanics and Engineering}, 2016.

\bibitem{Bourdin:2008a}
B.~Bourdin, G.~A. Francfort, and J.-J. Marigo.
\newblock The variational approach to fracture.
\newblock {\em Journal of Elasticity}, 91(1):5--148, 2008.

\bibitem{Bourdin:2014a}
B.~Bourdin, J.-J. Marigo, C.~Maurini, and P.~Sicsic.
\newblock Morphogenesis and propagation of complex cracks induced by thermal
  shocks.
\newblock {\em Phys. Rev. Lett.}, 112:014301, Jan 2014.

\bibitem{Chen:2017a}
Chih-Hung Chen, Eran Bouchbinder, and Alain Karma.
\newblock Instability in dynamic fracture and the failure of the classical
  theory of cracks.
\newblock {\em Nat Phys}, advance online publication:--, 08 2017.

\bibitem{Chen:2015b}
Chih-Hung Chen, Tristan Cambonie, Veronique Lazarus, Matteo Nicoli, Antonio~J.
  Pons, and Alain Karma.
\newblock Crack front segmentation and facet coarsening in mixed-mode fracture.
\newblock {\em Phys. Rev. Lett.}, 115:265503, Dec 2015.

\bibitem{Connally:1992}
J.~a. Connally and S.~B. Brown.
\newblock Slow crack growth in single-crystal silicon.
\newblock 256(5063):1537--1539, 1992.

\bibitem{Dauskarat:1990}
Reinhold~H Dauskarat, David~B Marshall, and Robert~O Ritchie.
\newblock Cyclic fatigue-crack propagation in magnesia-partially-stabilized
  zirconia ceramics.
\newblock {\em Journal of the American Ceramic Society}, 73(4):893--903, 1990.

\bibitem{Deshpande:2002}
V.~S. Deshpande, A.~Needleman, and E.~Van~der Giessen.
\newblock Discrete dislocation modeling of fatigue crack propagation.
\newblock {\em Acta Materialia}, 50(4):831--846, 2002.

\bibitem{Fender:2010}
Melissa~L Fender, Fr{\'e}d{\'e}ric Lechenault, and Karen~E Daniels.
\newblock Universal shapes formed by two interacting cracks.
\newblock {\em Physical review letters}, 105(12):125505, 2010.

\bibitem{Forman:1967}
R.~G. Forman, V.~E. Kearney, and R.~M. Engle.
\newblock Numerical analysis of crack propagation in cyclic-loaded structures.
\newblock {\em Journal of Basic Engineering}, 89(3):459--463, 09 1967.

\bibitem{Francfort:1998}
G.A. Francfort and J.J. Marigo.
\newblock {Revisiting brittle fracture as an energy minimization problem}.
\newblock {\em Journal of the Mechanics and Physics of Solids},
  46(8):1319--1342, 1998.

\bibitem{Ghelichi:2015}
Ramin Ghelichi and Ken Kamrin.
\newblock Modeling growth paths of interacting crack pairs in elastic media.
\newblock {\em Soft Matter}, 11(40):7995--8012, 2015.

\bibitem{Gravouil:2002}
A.~Gravouil, N.~Mo{\"e}s, and T.~Belytschko.
\newblock Non-planar 3d crack growth by the extended finite element and level
  sets---part ii: Level set update.
\newblock {\em International Journal for Numerical Methods in Engineering},
  53(11):2569--2586, 2002.

\bibitem{Gurson:1977a}
A.~L. Gurson.
\newblock Continuum theory of ductile rupture by void nucleation and growth:
  Part i---yield criteria and flow rules for porous ductile media.
\newblock {\em Journal of Engineering Materials and Technology}, 99(1):2--15,
  01 1977.

\bibitem{Hakim:2009}
Vincent Hakim and Alain Karma.
\newblock Laws of crack motion and phase-field models of fracture.
\newblock {\em Journal of the Mechanics and Physics of Solids}, 57(2):342--368,
  2 2009.

\bibitem{Hosseini:2018}
Zahra~S Hosseini, Mohsen Dadfarnia, Brian~P Somerday, Petros Sofronis, and
  Robert~O Ritchie.
\newblock On the theoretical modeling of fatigue crack growth.
\newblock {\em Journal of the Mechanics and Physics of Solids}, 121:341--362,
  2018.

\bibitem{Kahn:2004}
H~Kahn, R~Ballarini, and AH~Heuer.
\newblock Dynamic fatigue of silicon.
\newblock {\em Current Opinion in Solid State and Materials Science},
  8(1):71--76, 2004.

\bibitem{Karma:2001a}
Alain Karma, David~A. Kessler, and Herbert Levine.
\newblock Phase-field model of mode iii dynamic fracture.
\newblock {\em Phys. Rev. Lett.}, 87:045501, Jul 2001.

\bibitem{libmesh}
B.~S. Kirk, Peterson~J. W., Stogner~R. H., and Carey~G. F.
\newblock {\texttt{libMesh}: A C++ Library for Parallel Adaptive Mesh
  Refinement/Coarsening Simulations}.
\newblock {\em Engineering with Computers}, 22(3--4):237--254, 2006.

\bibitem{Koivisto:2016}
Juha Koivisto, M-J Dalbe, MJ~Alava, and S~Santucci.
\newblock Path (un) predictability of two interacting cracks in polycarbonate
  sheets using digital image correlation.
\newblock {\em Scientific reports}, 6, 2016.

\bibitem{Lange:1968}
FF~Lange.
\newblock Interaction between overlapping parallel cracks; a photoelastic
  study.
\newblock {\em International Journal of Fracture}, 4(3):287--294, 1968.

\bibitem{Legrand:2015}
L~Legrand and V~Lazarus.
\newblock Front shape and loading evolution during cracks coalescence using an
  incremental perturbation method.
\newblock {\em Engineering Fracture Mechanics}, 133:40--51, 2015.

\bibitem{lubomirsky2018universality}
Yuri Lubomirsky, Chih-Hung Chen, Alain Karma, and Eran Bouchbinder.
\newblock Universality and stability phase diagram of two-dimensional brittle
  fracture.
\newblock {\em Physical review letters}, 121(13):134301, 2018.

\bibitem{Melin:1983}
Solveig Melin.
\newblock {Why do cracks avoid each other?}
\newblock {\em International Journal of Fracture}, 23:37--45, 1983.
\newblock 10.1007/BF00020156.

\bibitem{Mesgarnejad:2013}
A.~Mesgarnejad, B.~Bourdin, and M.~M. Khonsari.
\newblock A variational approach to the fracture of brittle thin films subject
  to out-of-plane loading.
\newblock {\em Journal of mechanics and physics of solids}, 61(11):2360--2379,
  November 2013.

\bibitem{Mesgarnejad:2015}
A.~Mesgarnejad, B.~Bourdin, and M.~M. Khonsari.
\newblock Validation simulations for the variational approach to fracture
  mechanics.
\newblock {\em Computer Methods in Applied Mechanics and Engineering}, 2015.

\bibitem{Mozaffari:2015}
Navid Mozaffari and George~Z. Voyiadjis.
\newblock Phase field based nonlocal anisotropic damage mechanics model.
\newblock {\em Physica D: Nonlinear Phenomena}, 308:11--25, 7 2015.

\bibitem{Muhlstein:2001}
CL~Muhlstein, SB~Brown, and RO~Ritchie.
\newblock {High-cycle fatigue and durability of polycrystalline silicon thin
  films in ambient air}.
\newblock {\em Sensors and Actuators A: Physical}, 94(3):177--188, 2001.

\bibitem{Needleman:1984}
A.~Needleman and V.~Tvergaard.
\newblock An analysis of ductile rupture in notched bars.
\newblock {\em Journal of the Mechanics and Physics of Solids}, 32(6):461--490,
  1984.

\bibitem{Nguyen:2001}
O~Nguyen, EA~Repetto, Michael Ortiz, and RA~Radovitzky.
\newblock A cohesive model of fatigue crack growth.
\newblock {\em International Journal of Fracture}, 110(4):351--369, 2001.

\bibitem{Ogura:1985}
Keiji Ogura, Yoshio Miyoshi, and Izuru Nishikawa.
\newblock Fatigue crack growth and closure of small cracks at the notch root.
\newblock {\em Current research on fatigue cracks(A 86-33002 14-39). Kyoto,
  Society of Materials Science, 1985,}, pages 57--78, 1985.

\bibitem{Paris:1963}
P.~Paris and F.~Erdogan.
\newblock A critical analysis of crack propagation laws.
\newblock {\em Journal of Basic Engineering}, 85(4):528--533, 12 1963.

\bibitem{Paris:1961}
Paul~C Paris, Mario~P Gomez, and William~E Anderson.
\newblock A rational analytic theory of fatigue.
\newblock {\em The trend in engineering}, 13(1):9--14, 1961.

\bibitem{Peacock:2002}
D.~C.~P Peacock.
\newblock Propagation, interaction and linkage in normal fault systems.
\newblock {\em Earth-Science Reviews}, 58(1):121--142, 2002.

\bibitem{Peco:2019}
C.~Peco, Y.~Liu, C.~Rhea, and J.~E. Dolbow.
\newblock Models and simulations of surfactant-driven fracture in particle
  rafts.
\newblock {\em International Journal of Solids and Structures},
  156-157:194--209, 2019.

\bibitem{Pham:2013}
Kim Pham and Jean-Jacques Marigo.
\newblock From the onset of damage to rupture: construction of responses with
  damage localization for a general class of gradient damage models.
\newblock {\em Continuum Mechanics and Thermodynamics}, pages 1--25, 2013.

\bibitem{Rice:1967}
J~Rice.
\newblock Mechanics of crack tip deformation and extension by fatigue.
\newblock In {\em Fatigue crack propagation}. ASTM International, 1967.

\bibitem{Ritchie:1988}
RO~Ritchie.
\newblock Mechanisms of fatigue crack propagation in metals, ceramics and
  composites: role of crack tip shielding.
\newblock {\em Materials Science and Engineering: A}, 103(1):15--28, 1988.

\bibitem{Ritchie:2005}
RO~Ritchie.
\newblock Incomplete self-similarity and fatigue-crack growth.
\newblock {\em International Journal of Fracture}, 132(3):197--203, 2005.

\bibitem{Ritchie:1999}
Robert~O Ritchie.
\newblock Mechanisms of fatigue-crack propagation in ductile and brittle
  solids.
\newblock {\em International Journal of Fracture}, 100(1):55--83, 1999.

\bibitem{Ritchie:1991}
Robert~O Ritchie and Reinhold~H Dauskardt.
\newblock Cyclic fatigue of ceramics.
\newblock {\em Journal of the Ceramic Society of Japan}, 99(1154):1047--1062,
  1991.

\bibitem{Suresh:1998}
Subra Suresh.
\newblock {\em Fatigue of materials}.
\newblock Cambridge university press, 1998.

\bibitem{Tanne:2017}
Erwan Tann\'{e}, Tianyi Li, Blaise Bourdin, J.-J Marigo, and Corrado Maurini.
\newblock Crack nucleation in variational phase-field models of brittle
  fracture.
\newblock 2017.

\bibitem{Walker:1970}
K~Walker.
\newblock The effect of stress ratio during crack propagation and fatigue for
  2024-t3 and 7075-t6 aluminum.
\newblock In {\em Effects of environment and complex load history on fatigue
  life}. ASTM International, 1970.

\bibitem{Wesley-Patterson:2010}
G.~Wesley~Patterson and James~W. Head.
\newblock Segmented lineaments on europa: Implications for the formation of
  ridge complexes and bright bands.
\newblock {\em Icarus}, 205(2):528--539, 2010.

\bibitem{Wohler:1860}
A~W{\"o}hler.
\newblock Versuche {\"u}ber die festigkeit der eisenbahnwagenachsen.
\newblock {\em Zeitschrift f{\"u}r Bauwesen}, 10:160--161, 1860.

\bibitem{Xue:2008}
Liang Xue.
\newblock Constitutive modeling of void shearing effect in ductile fracture of
  porous materials.
\newblock {\em Engineering Fracture Mechanics}, 75(11):3343--3366, 2008.

\end{thebibliography}

\end{document}